\documentclass[structabstract]{aa} % for the abstract without structuration 
\usepackage{graphicx}
\usepackage{natbib}

\usepackage{txfonts}
\usepackage[cmyk]{xcolor}
\usepackage{tikz}

\begin{document}
\title{Dusty wind of W\,Hya.}
\subtitle{Multi-wavelength modelling of the present-day and recent mass loss}

   \author{T. Khouri
           \inst{1}\thanks{{\it Send offprint requests to T. Khouri}\newline \email{theokhouri@gmail.com}}, L. B. F. M. Waters \inst{1,2}, A. de Koter \inst{1,3}, L. Decin\inst{1,3},
           M. Min \inst{1}, B. L. de Vries \inst{4,5}, R. Lombaert \inst{3}, N. L. J. Cox \inst{3}}

\institute{Astronomical Institute ``Anton Pannekoek", University of Amsterdam, PO Box 94249, 1090 GE Amsterdam, The Netherlands %1
            \and
            SRON Netherlands Institute for Space Research, Sorbonnelaan 2, 3584 CA Utrecht, The Netherlands %2
            \and
            Instituut voor Sterrenkunde, KU Leuven, Celestijnenlaan 200D B-2401, 3001 Leuven, Belgium %3
            \and
            AlbaNova University Centre, Stockholm University, Department of Astronomy, 106 91 Stockholm, Sweden%4
	  \and
            Stockholm University Astrobiology Centre, 106 91 Stockholm, Sweden%5
}
   \date{}

 \authorrunning{Khouri et al.}
  \abstract
   {Low- and intermediate-mass stars go through a period of intense mass-loss at the end of their lives during the asymptotic giant branch (AGB) phase. While on the AGB a significant part, or even most, of their initial mass is expelled in
   a stellar wind.
   This process controls the final stages of the evolution of these stars and contributes to the chemical evolution of galaxies.
   However, the wind-driving mechanism of AGB stars is not yet well understood, especially so for oxygen-rich sources. Characterizing both the present-day mass-loss
   rate and wind structure and the evolution of the mass-loss rate of such stars is paramount to advancing our understanding of this processes.}
   {We study the dusty wind of the oxygen-rich AGB star W\,Hya to understand its composition and structure and shed light on the mass-loss mechanism.}
   {We modelled the dust envelope of W\,Hya using an advanced radiative transfer code. We analysed our dust model in the light of a previously calculated gas-phase wind
   model and compared it with measurements available in the literature, such as infrared spectra, infrared images, and optical scattered light fractions.}
   {We find that the dust spectrum of W\,Hya can partly be explained by a gravitationally bound dust shell that probably is responsible for most of the amorphous Al$_2$O$_3$ emission.
   The composition of the large ($\sim$\,0.3\,$\mu$m) grains needed to explain the scattered light cannot be constrained, but probably is dominated by silicates.
   Silicate emission in the thermal infrared was found to originate from beyond 40 AU from the star. In our model, the silicates need to have substantial near-infrared opacities to be visible at such
   large distances. The increase in near-infrared opacity of the dust at these distances roughly coincides with a sudden increase in expansion velocity as deduced from the gas-phase CO lines.
   The dust envelope of W\,Hya probably contains an important amount of calcium but we were not able to obtain a dust model that reproduces the observed emission while
   respecting the limit set by the gas mass-loss rate.
   Finally, the recent mass loss of W\,Hya is confirmed to be highly variable and we identify a strong peak in the mass-loss rate that occurred about 3500 years ago
   and lasted for a few hundred years.
   }
   {}


   \maketitle
%
%________________________________________________________________

\section{Introduction}

Low- and intermediate-mass stars in the last stages of their lives evolve to the asymptotic giant branch (AGB)
and occupy the region of the Hertzsprung-Russell diagram of high luminosities (typically 5000 -- 50000 L$_\odot$)
and low effective temperatures ($\sim$\,3000\,K). Dust grains can form in the outer layers of
the extended atmospheres of AGB stars where the temperatures are low and the densities are enhanced by stellar pulsations.
Radiation pressure acting on the grains causes a slow wind to develop ($\sim$\,10\,km/s) through which the star
looses mass, a process that controls the subsequent evolution of the star \citep{Habing2003}.
The cumulative mass lost during the AGB phase is important for the chemical evolution of galaxies,
as the atmospheric composition of AGB stars is modified by dredge-ups,
which bring elements synthesized in the stellar interior to the surface \citep[e.g.][]{Habing1996}.

Despite a qualitative understanding of the AGB mass-loss process, models are not yet able to predict the
mass-loss history of an AGB star from first principles. Both the magnitude of the mass loss and its variation in
time are still unanswered questions. For instance, model calculations show that the winds of oxygen-rich AGB stars cannot 
be driven by absorption of photons. The reason is that iron-bearing silicates that have large near-infrared absorption 
cross-sections to acquire enough momentum also heat up very efficiently and cannot exist in the wind acceleration region \citep{Woitke2006}.
A solution to this problem was proposed by \cite{Hofner2008}, 
who suggested that refractory species with small near-infrared absorption cross-sections -- that can exist close to the star -- can still transfer momentum 
to the wind by means of scattering, if the grains can grow to a few tenths of a micron in diameter.
Indeed, in a more comprehensive study of the possible wind-drivers in oxygen-rich AGB stars, \cite{Bladh2013} found that the dust
species expected to form in an oxygen-rich environment are not able to drive the wind via the absorption of photons.
The nature of the grains responsible for driving the wind and the particulars of the wind-driving
mechanism are still widely discussed \citep[e.g.][]{Norris2012,Bladh2012,Bladh2013,Gail2013,Karovicova2013}.
From wind-driving calculations, the preferred wind-driving candidates are silicate grains \citep{Hofner2008,Bladh2012}.
However, amorphous aluminum oxide is also
found to be an important component in general oxygen-rich AGB stars with low mass-loss rates
\citep[e.g.][]{Lorenz-Martins2000,Speck2000,Heras2005,Karovicova2013}. The latter might act as a seed for further grain growth and be crucial for
starting the wind, despite their low abundance.
Aluminum oxide has been suggested to form in the extended atmospheres of oxygen-rich late-type stars 
both on the basis of grain condensation calculations \citep[e.g.][]{Woitke2006} and observations 
\citep[e.g.][]{Verhoelst2006,Zhao-Geisler2011}.  MIDI visibilities of RT\,Vir, for instance, point to the
presence of amorphous Al$_{2}$O$_{3}$ close to the star (at $\sim$\,1.5\,R$_\star$; \citeauthor{Sacuto2013}
\citeyear{Sacuto2013}).  However, dust excess spectral fits often require a too large amount of this grain species to match
the observed infrared emission; so large that it often implies a super-solar aluminum abundance in the
outflow \citep[e.g.][]{Karovicova2013}.

We model the dust envelope of W\,Hya, a close-by oxygen-rich AGB star with a fairly low mass-loss rate \citep[ henceforth Paper I]{Khouri2014} for which a wealth of observational data is available.
These data probe the wind structure of W\,Hya on spatial scales ranging from tens of mas to minutes of arc and reveal
grains with radii of about 0.3 $\mu$m at 40 mas (or 2\,R$_{\star}$) seen in scattered light \citep{Norris2012};
an inner radius for the amorphous silicate emission of 500 mas (or 25\,R$_{\star}$; \citeauthor{Zhao-Geisler2011}
\citeyear{Zhao-Geisler2011}), and a mass loss that is variable on dynamical timescales of thousands of years
\citep{Hawkins1990,Cox2012} that resembles the formation of detached-shells (exclusively) seen around carbon-rich AGB stars
\citep[e.g.][]{Olofsson1990,Olofsson1996}.  Although W\,Hya has previously been studied extensively, the models obtained for the
dust excess only focused
on the emission from the inner wind \citep[e.g.][]{Heras2005,Justtanont2004,Khouri2014}.  None have accounted
for the scattered light fractions seen by \citeauthor {Norris2012} or the inner emission radius of
the amorphous silicates reported by \citeauthor{Zhao-Geisler2011}, however.  Furthermore, all dust emission models obtained so far
require super-solar aluminum abundances in the dust when compared to the gas-phase mass-loss rate.

Based on the strong evidence for amorphous Al$_2$O$_3$ emission coming from very close to the star and the super-solar abundances in the wind that are supposedly  needed
to account for the amorphous Al$_2$O$_3$ emission, we consider in our model a gravitationally bound shell of amorphous Al$_2$O$_3$ grains to be
the main contributor to the emission at wavelengths around 12\,$\mu$m. We combine this gravitationally bound shell with an outflow to obtain a dust model
that fits the wealth of information on the dusty wind of W\,Hya.   The properties and spatial
distribution of the dust in this model are discussed in the context of wind driving, as well as the time variability of the
mass-loss on timescales of millennia.

In Sect. \ref{sec:dust_obs} we detail our modelling strategy and the observations of the dust envelope of W\,Hya.
The modelling procedure and fits are described in Sect. \ref{sec:new_model}. We discuss
our results in Sect. \ref{sec:disc} and provide a summary in Sect. \ref{sec:summary}.

\section{Observations of the dust envelope of W\,Hya}
\label{sec:dust_obs}

In the following subsections, we present the dataset used in our study to constrain the present-day mass-loss rate.
The observations used to model the recent dust mass-loss history are presented in Sect. \ref{sec:pacs_image}.

\begin{table*}
\caption{Overview of the observations of the dust shell of W\,Hya using different telescopes. The phase $\Phi$ of the pulsation period is given based on observations
of visible light made available by the AAVSO$^{\rm a}$}.            
\label{tab:dust_obs} 
\centering                          
\begin{tabular}{l c @{\hspace{0.2cm}}c l @{\hspace{0.2cm}}c @{\hspace{0.2cm}}c r}        
\hline\hline                 
Instrument & Date & $\Phi$ & Ref. & $\lambda$ [$\mu$m] & FOV [$'' \times '' $]& Identifier \ \\    
\hline                        
ISO SWS 1 & 14-02-1996 & 0.12 & \cite{Sloan2003} & 2.36 -- 45.39 & $14\times20$ to $17\times40$& 08902004 \\
ISO SWS 2 & 07-01-1997 & 0.17 & \cite{Justtanont2004} & 2.36 -- 45.39 & $14\times20$ to $17\times40$ & 41800303 \\
ISO LWS 1 & 07-02-1996 & 0.1 & ISO data archive$^{\rm b}$ & 43.05 -- 195.5 & $80\times80$ to $70\times70$ & 08200208 \\
ISO LWS 2 & 05-02-1997 & 0.1 & ISO data archive$^{\rm b}$ & 43.05 -- 195.5 & $80\times80$ to $70\times70$ & 44700672 \\
ISO LWS 3 & 02-08-1997 & 0.6 & ISO data archive$^{\rm b}$ & 43.05 -- 195.5 & $80\times80$ to $70\times70$ & 62500572 \\
PACS Spec B2A & 09-07-2011 & 0.8 & \cite{Khouri2014} & 55 -- 73 & $30\times30$ & 1342212604 \\
PACS Spec R1A & 09-07-2011 & 0.8 & \cite{Khouri2014} & 102 -- 146  & $30\times30$ & 1342212604 \\
PACS Spec B2B & 14-01-2011 & 0.4 & \cite{Khouri2014} & 70 -- 105 & $30\times30$ & 1342223808 \\
PACS Spec R1B & 14-01-2011 & 0.4 & \cite{Khouri2014} & 140 -- 210 & $30\times30$ & 1342223808 \\
SPIRE Spec & 09-01-2010 & 0.27 & \cite{Khouri2014} & 124 -- 671 & $37\times37$ to $18\times18$ & 1342189116 \\
PACS Image 70 & 08-02-2011 & 0.47 & \cite{Cox2012} & 60 -- 85 & $1440\times1440$ & 1342213848 \\
PACS Image 160 & 08-02-2011 & 0.47 & \cite{Cox2012} & 130 -- 210 & $1440\times1440$ & 1342213849 \\
SCUBA & & & & 450 and 850\\
VLT - MIDI & 04-2007 -- 09-2009 & - & \cite{Zhao-Geisler2011} & 8 -- 13 \\
VLT - NACO & 03-2009 and 06-2010 & 0.2$^{\rm c}$ & \cite{Norris2012} & 1.04, 1.24 and 2.06 \\
\hline                                   
\end{tabular}
\tablefoot{ a - American Association of Variable Star Observers at http://www.aavso.org/, b - http://iso.esac.esa.int/iso/ida/,
 c - Visual phase given by \cite{Norris2012}.}
\end{table*}

\subsection{Observed infrared spectra}

\begin{figure}[h]
   \centering
   \includegraphics[width= 8.5cm]{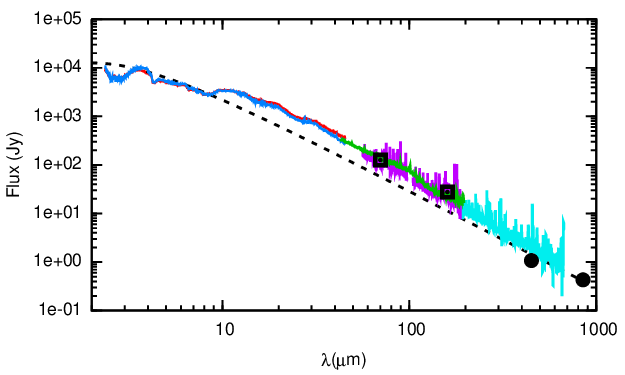}
      \caption{Compilation of all thermal emission observations of the dust in W\,Hya. The red and blue solid lines are the ISO SWS spectra from 1996 and 1997, respectively,
      the green solid line is the averaged ISO LWS spectrum, the purple and light blue solid lines are the PACS and SPIRE spectra, and the black squares and black circles depict the PACS
      and SCUBA photometric measurements. A black-body spectrum of 2500\,K and 5400\,L$_\odot$ is overplotted to guide the eye (black dashed line).}
         \label{fig:obs_spec}
   \end{figure}

W\,Hya was observed by the short-wavelength spectrometer \citep[SWS,][]{deGraauw1996} and the long-wavelength spectrometer \citep[LWS,][]{Clegg1996}
onboard the infrared space observatory \cite[ISO,][]{Kessler1996} and by the photodetector array camera and spectrometer \citep[PACS, ][]{Poglitsch2010} and the spectral and photometric imaging receiver
\citep[SPIRE,][]{Griffin2010} onboard {\em Herschel} Space Observatory \citep{Pilbratt2010}.
A spectrum covering the full range of the instruments was obtained with both PACS and SPIRE.
The PACS observations were carried out in different epochs for the blue and red band. The full
spectral range of SPIRE was covered in one observation run (see also Table \ref{tab:dust_obs}). Two spectra
were measured with SWS and three with LWS. The two SWS spectra were taken almost one pulsation period
apart (Table \ref{tab:dust_obs}). We obtained the reduced 1997 spectrum from \cite{Justtanont2004} and the 1996 spectrum from
the \cite{Sloan2003} database.
Two LWS observations were also taken roughly one period apart, with a third taken
near maximum visual light. We obtained the highly processed data products of the three spectra from the ISO archive,
in which the problem of near-infrared leakage is corrected. The
spectrum obtained in February 1997 has an unreliable baseline, especially at long wavelengths,
and was not included in our analysis. The other two LWS spectra show very similar flux levels
(displaying differences of about 10\%) and shapes. We have averaged these two spectra before fitting our models.
All the observed spectra are shown in Fig. \ref{fig:obs_spec}; the
two SWS spectra differ slightly in flux level, revealing differences of up to 20\%,
but are very similar in shape. The PACS and SPIRE flux levels agree
well with those of LWS when the different fields of view and the PACS far-infrared maps
are taken into account (see Sect. \ref{sec:new_model}).

\begin{figure}[h]
   \centering
   \includegraphics[width= 8.5cm]{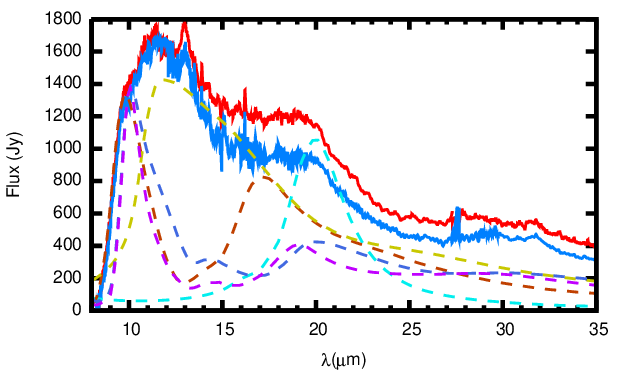}
      \caption{Continuum-subtracted ISO SWS spectra from 1996 and 1997 are shown by the red and blue lines.
      The opacity curves of amorphous Al$_2$O$_3$, MgFeSiO$_4$, FeO, Ca$_2$Mg$_{0.5}$Al$_2$Si$_{1.5}$O$_7$ and Ca$_2$Al$_2$SiO$_7$
      are shown by the yellow, light brown, light blue, purple, and blue dashed lines, respectively. The opacity curves are scaled to compared them with the
      stellar-continuum-subtracted spectrum.}
         \label{fig:comp_opas}
   \end{figure}

The dust components seen in the ISO spectrum of W\,Hya are better visualized if we subtract the expected
stellar continuum from the observed spectrum. We used the stellar parameters and distance from Paper I, T$_{\rm eff}$ = 2500\,K,
L=5400\,L$_\odot$ and 78\,pc, to estimate the stellar continuum.
In Fig. \ref{fig:comp_opas}, we compare the opacity curves of the dust species considered by
\cite{Cami2002} and \cite{Heras2005} and that of Ca$_2$Mg$_{0.5}$Al$_2$Si$_{1.5}$O$_7$ to the two
stellar-continuum-subtracted SWS spectra. A combination of amorphous silicates,
amorphous Al$_2$O$_3$, and Mg$_{0.1}$Fe$_{0.9}$O can qualitatively account for the emission seen between
9\,$\mu$m and 25\,$\mu$m. From $\sim$\,27\,$\mu$m to $\sim\,32\,\mu$m, however, a feature is observed that cannot be reproduced by
these dust species. We refer to this feature as the broad 30\,$\mu$m feature.
\cite{Fabian2001} identified a narrow feature at 32\,$\mu$m that they assigned to
spinel, but this is different from the broad 30\,$\mu$m feature we refer to here.

\subsection{Interferometric observations}
\label{sec:interf_obs}

The envelope of W\,Hya was studied with the Masked Aperture-Plane Interference Telescope (MAPPIT) at
the 3.9 m Anglo-Australian Telescope over the wavelength range 0.65 -- 1.0$\,\mu$m by \cite{Ireland2004}.
The authors found an increase in measured angular size from $\sim$\,20\,mas at 1$\,\mu$m to $\sim$\,35\,mas at shorter wavelengths, which they proposed
to be due the scattering of light by dust grains.
Later observations of W\,Hya by \cite{Norris2012} using aperture-masked polarimetric interferometry with the NACO instrument on the VLT
confirmed the existence of a shell of scattering dust particles around the star with a radius of
about 37.5 mas, or $\sim$\,2\,R$_\star$. The authors considered a population of large grains
with uniform radii in their model, as their technique does not allow them to constrain the presence of small grains, and found the scattering particles to be $0.316\pm 0.004\,\mu$m.
Although the composition of the dust could not be determined,
the authors suggested that they may be composed
of iron-free silicates such as forsterite (Mg$_2$SiO$_4$) or enstatite (MgSiO$_3$), while corundum (Al$_2$O$_3$) is a possible candidate as well.

\cite{Zhao-Geisler2011} monitored W\,Hya using MIDI at the VLT.  The authors found that the silicate emission
seen at 10 $\mu$m in the ISO spectrum must come from a region with an inner radius larger than 500 mas, corresponding to 25 times the stellar radius
or 12 times the scattering shell radius as estimated by \cite{Norris2012}. This does not exclude the possibility that silicates condense close to the
star, as the iron-free silicates proposed by \cite{Norris2012} to be the scattering agent are transparent in the near infrared
and can remain cold ($\sim$\,700\,K) and invisible even at two stellar radii.
Such grains would not necessarily produce a signature in the thermal infrared (see Sect. \ref{sec:GBDS}).

\subsection{Gas-phase wind model}

In Paper I, the present-day gas mass-loss rate of W\,Hya was determined to be $(1.3 \pm 0.6) \times 10^{-7}\, {\rm M}_\odot$\,year$^{-1}$
by fitting the strengths and shapes of the CO ground-vibrational state, pure-rotational transitions observed by {\em Herschel}.
This mass-loss rate places constraints on the amount of dust that can form in the wind assuming a solar composition.
More stringent limits on the amount of silicates that may form around W\,Hya were set by \cite{Khouri2014a}, who modelled the SiO gas-phase emission
lines observed by {\em Herschel} using the wind model obtained in Paper I. The authors found that about 60\% of the silicon in the outflow of W\,Hya remains in the gas phase.

\section{Towards a comprehensive present-day mass-loss model}
\label{sec:new_model}

Our modelling strategy consists of dealing with the different spatial and compositional components of the wind in an independent way first. This approach can be followed as
the wind of W\,Hya is optically thin at the wavelengths where dust emission is important. We start by modelling the innermost observed dust component,
namely the scattered light fractions and the amorphous Al$_2$O$_3$ emission. Then, we model the silicate emission envelope, whose inner radius is at ten times
the distance from where the scattered light is seen \citep{Zhao-Geisler2011}. Finally, we compare our best inner wind model with the PACS $70\,\mu$m image,
which provides constraints on both the dust distribution close to the star and at scales up to a few arcminutes \citep{Cox2012}.
In our modelling efforts we aim to reproduce the following observations:

\begin{itemize}
\item{The scattered light fractions as measured by \cite{Norris2012}.}
\item{The ISO dust spectrum.}
\item{The lower limit of 500 mas on the inner radius of the silicate emission shell set by \cite{Zhao-Geisler2011}.}
\item{The limits set by the gas-phase model of Paper I and \cite{Khouri2014a} on the amount of elements available for dust formation in the wind.}
\item{The broadness of the 70\,$\mu$m radial brightness distribution seen by PACS \citep{Cox2012}.}
\end{itemize}

\subsection{MCMax and model assumptions}
\label{sec:MCMax}

To solve the continuum radiative transfer in the wind of W\,Hya, we used the code {\sc MCMax} \citep{Min2009}. The code calculates the dust spectrum,
images of the envelope at different wavelengths, and the fraction of light scattered by the dust particles, all of which are used to constrain our model.
The dust envelope was assumed to be spherically symmetric.
To derive dust opacities from optical constants, we applied a distribution of hollow spheres \citep[DHS,][]{Min2003} approximation to represent the particles shapes.
We used the direction-dependent scattering phase-function in our calculations, since particles with radii in the range considered scatter light non-isotropically.
This was done by considering the full angle-dependent Mueller matrix for each scattering event.
The temperature of the dust grains was set by assuming radiative equilibrium. For multiple dust species it may be computed
assuming the different types of grains to be either in isolation or in thermal contact. In the case of thermal contact, the absorbed
energy is distributed over the constituents of the grain and a single grain temperature is computed such that the emitted energy
matches the absorbed energy. 
We specify throughout the text whenever thermal contact is considered.

To model the amorphous Al$_2$O$_3$ emission, we used three sets of optical constants, those for porous and compact particles given by \cite{Begemann1997} and those of 
aerosil particles measured by \cite{Koike1995}. The latter measurements were carried out for wavelengths between 0.5 and 500\,$\mu$m.
The optical constants obtained by \cite{Begemann1997} were measured between 7.7 and 500\,$\mu$m and 
lack data at near-infrared and optical wavelengths. As these short wavelength data are important for determining the temperature of the grains,
we used the optical constants given by \cite{Koike1995} at  $\lambda < 7.7\,\mu$m.
To model the silicate emission, we retrieved data for silicate dust species from the JENA database
of optical constants from the works of \cite{Jager1994,Dorschner1995,Mutschke1998}, and \cite{Jager2003}.
The dust species that are considered are given in Table \ref{tab:optConst}.
For many of the silicate species the optical constants are available only down to about 6 $\mu$m. As most of the radiation that heats the grains is absorbed
at shorter wavelengths, we considered the near-infrared optical constants of these dust species
to be equal to that of Mg$_{0.4}$Fe$_{0.6}$SiO$_3$, Mg$_{0.8}$Fe$_{0.2}$SiO$_3$ or MgSiO$_3$, selecting out of these three the species that best
matches the iron content.

We considered W\,Hya to be at 78 parsecs, to have a luminosity of 5400 L$_\odot$ and the stellar spectrum to be that of a black-body with 2500\,K,
following our assumptions in Paper I.

\subsection{Previous dust model}
\label{sec:prev_model}

The model obtained in Paper I for the wind of W\,Hya (shown in Fig. \ref{fig:newDustFit}) adopted the dust components identified by
\cite{Justtanont2004}. In Paper I, the region of the ISO spectrum between 8 and 30 $\mu$m of the ISO spectrum was fitted,
but no attempt was made to fit the 13 $\mu$m feature as its carrier is still uncertain
\citep[see][and references therein]{Posch1999,Zeidler2013}.
A total dust mass-loss rate of $2.8 \times 10^{-10}$ M$_{\odot}$\,year$^{-1}$ was obtained.
The dust composition by mass is 58 \% astronomical silicates \citep{Justtanont1992},
34 \% amorphous aluminum oxide (Al$_2$O$_3$), and 8 \% magnesium-iron oxide (MgFeO).
Optical constants for the latter species were retrieved from the database of the University of Jena from the work of \cite{Henning1995}.
For amorphous aluminum oxide, the authors have used data for porous particles from \cite{Begemann1997}.

\subsection{Amorphous Al$_2$O$_3$ emission}

\begin{figure}[t]
   \centering
   \includegraphics[width= 8.5cm]{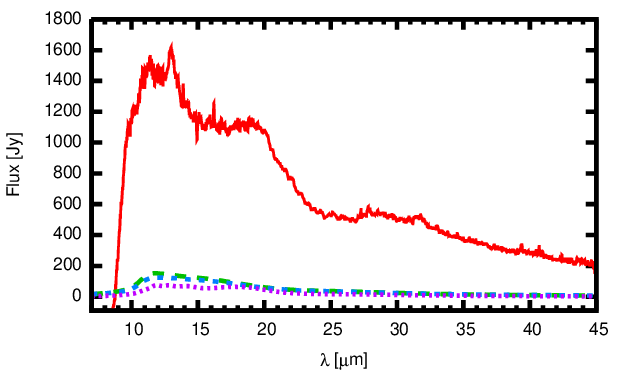}
      \caption{Models considering full condensation of aluminum in a solar-composition outflow compared to the stellar-continuum-subtracted infrared excess (solid red).
The aerosil Al$_2$O$_3$ particles, porous Al$_2$O$_3$ particles and,
      compact Al$_2$O$_3$ particles are represented by the dotted purple, short dashed blue and long dashed green lines. None of the considered
      species produces strong enough emission at 12\,$\mu$m.}
         \label{fig:Al2O3-wind}
   \end{figure}

As pointed out in the introduction, the amorphous Al$_2$O$_3$ mass fraction of the total dust needed to explain the mid-infrared
excess of low mass-loss rate AGB stars is often very high \citep[e.g.][]{Lorenz-Martins2000,Speck2000,Heras2005,Karovicova2013}.
The maximum amount of a given dust species that can exist in the stellar wind is determined by the availability
of its least abundant element. In the case of amorphous Al$_2$O$_3$ this limit is set by aluminum, whose
mass abundance relative to hydrogen is $7.6\times10^{-5}$ for a solar composition \citep{Asplund2009}.
If all aluminum atoms in the wind of W\,Hya were used to form amorphous Al$_2$O$_3$ grains, the mass-loss rate of this dust species alone would be
${\rm \dot M_{\rm Al_2O_3}} = 1.3\times10^{-7} \times 7.6\times10^{-5} / 0.53= 1.9\times10^{-11}$\,M$_\odot$ year$^{-1}$, where
0.53 is the fraction of the mass of amorphous Al$_2$O$_3$ in aluminum atoms and $1.3\times10^{-7}$\,M$_\odot$\,year$^{-1}$ the gas mass-loss rate.
Assuming the specific weight of amorphous Al$_2$O$_3$ to be 4.0 g cm$^{-3}$, we computed models with this amorphous Al$_2$O$_3$
mass loss, condensing at two stellar radii. At 12\,$\mu$m, the excess emission produced by these models is only 5\,\% of what is seen in the ISO spectrum (see Fig. \ref{fig:Al2O3-wind}).
These calculations depend on the adopted grain model, the optical constants used, and the assumed location of the onset of the wind.
We investigated these parameters using the wind model derived in Paper I by modifying the opacities for amorphous Al$_2$O$_3$
at short wavelengths and varying the inner radius of the dust envelope accordingly. We find that even an increase in the near-IR opacity
of amorphous Al$_2$O$_3$ by a factor of 10 increases the emission at 12\,$\mu$m by only about a factor of four. Increasing the near-IR opacity
also leads to an important contribution of amorphous Al$_2$O$_3$ emission at wavelengths shorter than 10\,$\mu$m. We conclude that
the emission attributed to amorphous Al$_2$O$_3$ in W\,Hya cannot be accounted for by such grains in the present-day wind alone.

\subsection{Gravitationally bound dust shell}
\label{sec:GBDS}

We calculated models for a gravitationally bound dust shell (GBDS) located at about 2.0\,R$_\star$ to account for the amorphous Al$_2$O$_3$ emission and the scattered light fractions.
The GBDS is composed of grains located in the region between the radius of onset of condensation of amorphous Al$_2$O$_3$ and that at which the radiation
pressure on the dust grains overcomes the pull of gravity. Since we do not calculate the dust condensation nor the transfer of momentum to the dust grains,
the inner and outer radii of this region are free parameters in our models.
In the GBDS the grains must reach the optimal size for scattering, at which radiation forces can accelerate them.
We therefore expect to find a distribution of sizes ranging from small grains,
with radii of the order of $0.01\,\mu$m to larger grains with radii of about $0.3\,\mu$m.
As an instructive exercise, we can approximate the size distribution by a mix of two particle sizes, small and large grains.
These can be studied somewhat independently based on the observations available:
large grains dominate the scattering of radiation, and small grains do not contribute to the scattered light and probably dominate the infrared emission.
We have considered small grains of $0.03\,\mu$m and large grains of $0.3\,\mu$m.

We considered the grains in the GBDS to be composed of either amorphous Al$_{2}$O$_{3}$ or Mg$_{2}$SiO$_{4}$. The amorphous Al$_{2}$O$_{3}$ emission can be fitted equally
well with large and small grains of this dust species.   Silicate emission from close to the star is not detected (Zhao-Geisler et al. 2011) but this does not necessarily mean
that silicates are absent.  Iron-free silicates (such as Mg$_{2}$SiO$_{4}$) can exist at close distances from the surface because they are translucent in the optical and near-infrared
and therefore remain relatively cold. This does imply that we will not be able to constrain the mass in small Mg$_{2}$SiO$_{4}$ grains with our models.
A fit to the observed scattered light fractions requires large grains and in principle can be obtained with either type of grain. Such a fit sets an upper limit on the amount
of large grains of the given species.

We considered the amorphous Al$_{2}$O$_{3}$ emission to originate from a stationary shell extending from 1.7 to 2.0 R$_\star$. The inner radius was set at that at which the
small amorphous Al$_2$O$_3$ grains reach temperatures of 1400\,K. The outer radius was assumed to be that found by \cite{Norris2012} for the origin of the scattered light.
In principle the grains in this region may be created and destroyed as material rises and falls as a consequence of stellar pulsations, but that does not affect our results.
Our modelling shows that the amount of emission is only mildly sensitive to the exact value of the boundaries or to a radial density gradient within the shell, but mostly to
the total mass of emitting amorphous Al$_{2}$O$_{3}$. The scattered light fractions were fitted considering a shell with an inner radius of 2.0 R$_\star$\ \citep{Norris2012} and
thickness of 0.1 R$_\star$, that is, enclosing the shell from where amorphous Al$_{2}$O$_{3}$ originates. In the context of our model, this shell of large grains
may be understood as being the interface between the GBDS and the outflow.

A population of only small amorphous Al$_2$O$_3$ grains reaches temperatures of 1250\,K at 2.0\,R$_\star$, while large
amorphous Al$_2$O$_3$ grains reach 1500\,K at the same distance.
The higher temperature of large amorphous Al$_2$O$_3$ grains has two causes.
First, the mass-absorption coefficient between 0.2 and 2$\,\mu$m is higher for the large grains than for small grains.
Consequently, large grains absorb more radiation by mass than their smaller counterparts and reach higher temperatures.
Second, the scattering of photons by large grains contributes to the diffuse radiation field and
the density of short-wavelengths photons ($\lambda \lesssim 2\,\mu$m) significantly increases.
The stronger radiation field at shorter wavelengths leads to more energy absorbed per time per grain and to a higher dust temperature.
This second effect also causes small amorphous Al$_2$O$_3$ particles to have higher temperatures when placed in a region with 0.3\,$\mu$m grains.
Assuming solar composition and full aluminum condensation, a shell of small amorphous Al$_2$O$_3$ grains between 1.7 and 2.0\,R$_\star$
requires a gas mass of $\sim$\,$1.4\times10^{-5}$\,M$_\odot$ to reproduce the amorphous Al$_2$O$_3$ emission seen in the ISO spectrum. Such a shell produces an infrared
spectrum indistinguishable from the amorphous Al$_2$O$_3$ models shown in Fig. \ref{fig:scatSpec} for a population with a range in grain radii.

The mass in large grains needed to reproduce the scattering is $4.9\times10^{-10}$ M$_\odot$ for amorphous Al$_2$O$_3$ and $3.6\times10^{-10}$ M$_\odot$ for Mg$_2$SiO$_4$ grains,
assuming specific weights of 4 g cm$^{-3}$ and 3.2 g cm$^{-3}$, respectively.
By assuming a solar composition, full aluminum condensation, and 35\% of silicon condensation \citep[as to be consistent with the available silicon for
dust condensation derived from the gas-phase analysis;][]{Khouri2014a}, we obtain gas masses of
$3.5\times10^{-6}$\,M$_\odot$ and $2.3\times10^{-7}$\,M$_\odot$ for large grains of amorphous Al$_2$O$_3$ and Mg$_2$SiO$_4$, respectively.

\begin{figure}[t]
   \centering
   \includegraphics[width= 8.5cm]{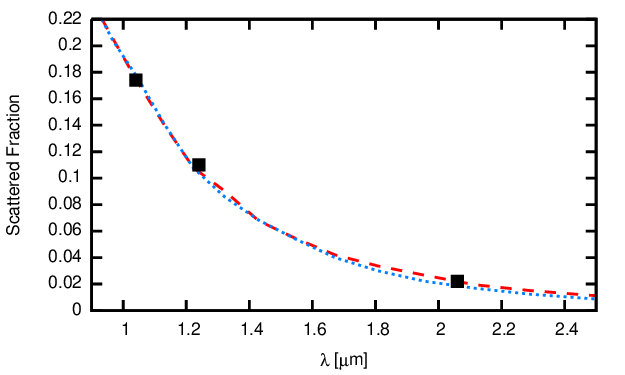}
      \caption{Scattering fractions for the models with large particles of aerosil Al$_2$O$_3$ (red long dashed line) and Mg$_2$SiO$_4$ (blue short dashed line) in the GBDS.
      The black squares are the observations by \cite{Norris2012}.}
         \label{fig:scatFrac}
   \end{figure}

\begin{figure}[h]
   \centering
   \includegraphics[width= 8.5cm]{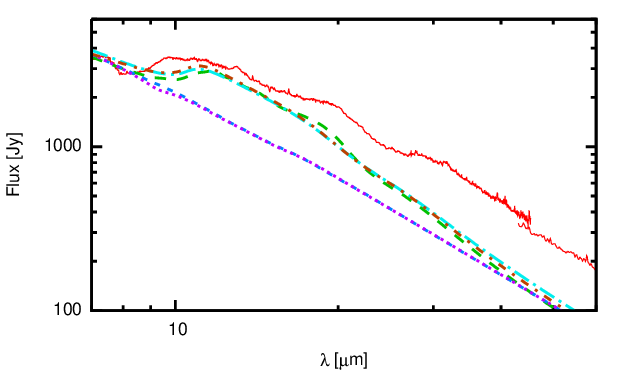}
      \caption{Spectra of a population of amorphous Al$_2$O$_3$, MgSiO$_3$ or Mg$_2$SiO$_4$ grains compared to the observed ISO spectrum \citep{Sloan2003}.
      We considered an MRN grain radii distribution for all models shown here.
      	The different lines are for grains consisting of aerosil Al$_2$O$_3$ particles (green dashed), Al$_2$O$_3$ porous particles (light blue long dashed-dotted),
      Al$_2$O$_3$ compact particles (brown short dashed-dotted), MgSiO$_3$ (purple dotted) and Mg$_2$SiO$_4$ (short dashed blue).}
         \label{fig:scatSpec}
   \end{figure}

If the fit to both the ISO infrared excess and the scattered light fractions are combined, the amorphous Al$_2$O$_3$ emission and the scattered light fractions can be explained by a single population of amorphous
Al$_2$O$_3$ grains located in a shell with inner and outer radii of 1.7 and 2.1 R$_\star$, respectively, and with a grain size distribution of the type, $n(a) \propto a^{-3.5}$,
where $a$ is the grain radius. This expression corresponds to the standard distribution for grain sizes found in the interstellar medium by \citeauthor{Mathis1977} (\citeyear{Mathis1977}, MRN distribution).
The lower limit of the distribution was assumed to be 0.01\,$\mu$m and the maximum grain radius to be that derived by \citeauthor{Norris2012}, i.e. 0.316\,$\mu$m.
By assuming solar composition and full aluminum condensation, the total gas mass required for this population of amorphous Al$_2$O$_3$ is $\sim$\,$1.2\times10^{-5}$\,M$_\odot$.
The differences in the amorphous Al$_2$O$_3$ mass needed to reproduce the observed emission
only considering small grains or a grain size distribution is due to the different grain temperatures obtained.
The fit to the scattered light fractions using an MRN size distribution is very similar to the one obtained with the large-grains-only model, shown in Fig. \ref{fig:scatFrac},
and is not given.
Gradients in the grain size throughout the GBDS may be present and would allow similar quality fits as adopting a size distribution. Infrared spectra for an MRN distribution
of particles adopting different optical constants for amorphous Al$_{2}$O$_{3}$ are shown in Fig. 5. The similar amorphous Al$_2$O$_3$ emission obtained
shows that the applied optical constants do not impact our findings significantly either.

If the scattering is due to large Mg$_2$SiO$_4$ grains, a population of small amorphous Al$_2$O$_3$ grains is still required to account for the infrared excess
emission. When a population of large silicate grains is placed in the shell together with small amorphous Al$_2$O$_3$ particles,
the temperatures are about 700\,K and 1400\,K, respectively.
When these two species are placed in thermal contact, the resulting temperature will depend on the mass fraction of each species in a given grain.
Assuming full aluminum and 35\% silicon condensation implies that 8\% of the mass is in amorphous Al$_2$O$_3$ and 92\% in Mg$_2$SiO$_4$. In this case,
the equilibrium temperature is about 800\,K and the silicate features would remain undetected, in agreement
with the limit set for the inner radius of amorphous silicate emission of $\sim$\,40\,AU.
Therefore, a scenario in which small amorphous Al$_2$O$_3$ grains are formed close to the star and in which rapid condensation of Mg$_2$SiO$_4$
onto the amorphous Al$_2$O$_3$ seeds occurs as well would be consistent with all observations. In this scenario, the condensation of Mg$_2$SiO$_4$
could be halted by the onset of the wind.

\subsection{Dust properties in the outflow}
\label{sec:scenario2}

\begin{figure}[t]
   \centering
   \includegraphics[width= 8.5cm]{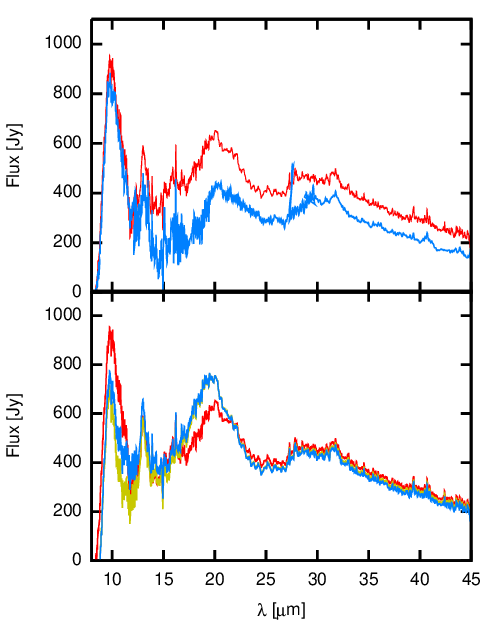}
      \caption{Al$_2$O$_3$-subtracted residual infrared spectrum of W\,Hya.
      {\it Upper panel:} Residuals adopting the optical constants for aerosil particles from \cite{Koike1995} and the ISO spectra from the \cite{Sloan2003} database (in red)
      and \cite{Justtanont2004} (in blue);
      {\it lower panel:} The red curve is the same as in the upper panel. The yellow and blue curves are for residuals assuming compact and porous particles from \cite{Begemann1997}.
      }
         \label{fig:sub-Al2O3Shell}
   \end{figure}

The contribution of amorphous Al$_2$O$_3$ grains in a GBDS to the infrared spectrum is somewhat independent of
the size distribution and the exact location of the shell as the grains would be sufficiently warm for the shape of their emission feature to be
independent of temperature for the optical constant adopted by us. Although the opacities of crystalline Al$_2$O$_3$ show significant
variations with temperature \citep{Zeidler2013}, to our knowledge, no studies have been performed that probe the effect of temperature variations on the amorphous Al$_2$O$_3$ features.
In the context of our model, only the strength of the emission is sensitive to temperature, and therefore the total grain mass needed to reproduce
the infrared excess is a function of grain temperature.
If we assume that the dust shell is optically thin, we may subtract the amorphous Al$_2$O$_3$ contribution from the ISO spectrum and study the residual spectrum (see Fig. \ref{fig:sub-Al2O3Shell}).
The assumed size distribution of amorphous Al$_2$O$_3$ grains does not affect the overall shape of the residual spectrum.
Uncertainties in the residual spectrum arise from the choice of optical constants for amorphous Al$_2$O$_3$,
the adopted grain model, and variability of the star, but these probably have a modest effect.

The residual spectrum clearly shows the silicate peak at 9.7 $\mu$m, the unidentified 13 $\mu$m feature, a peak at about 20 $\mu$m
that is probably due to 18 $\mu$m silicate and/or magnesium-iron oxide emission,
and a broad feature around 30 $\mu$m.
For simplicity and because of the similarity in the different Al$_2$O$_3$-subtracted residual spectra, we only used the ISO spectrum retrieved from the \cite{Sloan2003} database
subtracted by the spectrum of aerosil amorphous Al$_2$O$_3$ grains (Fig. \ref{fig:sub-Al2O3Shell}; red solid line).

\subsubsection{Fit to the residual spectrum}

The dust species considered in fitting the silicate emission are given in Table \ref{tab:optConst}.
All the species listed have somewhat similar opacity profiles that are characteristic of silicate glasses. Diagnostic parameters
such as the peak strength ratio, the central wavelength, and the full-width at half maximum (FWHM)
of the peaks  can help constrain the composition \citep{Dorschner1995,Mutschke1998} and are listed for DHS particles with 0.3\,$\mu$m radius.
The position of the two silicate peaks measured from the Al$_2$O$_3$-subtracted residual spectra are
$9.94\pm0.5\,\mu$m and $19.7\pm0.5$ $\mu$m. It is important to note that the peak at 19.7 $\mu$m might be produced by a combination of silicate and
Mg$_{\rm x}$Fe$_{\rm 1-x}$O emission, so the measured position should be considered with care.

We recall that the VLT 10\,$\mu$m observations provide a lower limit to the radial distance of the silicate-emitting shell of about 40 AU.
Most of the dust species listed in Table \ref{tab:optConst} have a low iron content and consequently a low near-infrared opacity. These species
would have too low temperatures at such large distances and would not produce detectable infrared flux relative to the bright stellar continuum and emission from the other dust species.
As the ratio between the two silicate emission peaks indicates that the grains must be warm ($\sim$\,500\,K),
the silicates with low near-infrared opacities, that is, with low iron content, were placed in thermal contact with metallic iron to increase their temperatures.
Such metallic iron inclusions are indeed expected to form in circumstellar silicates grains \citep{Gail1999}.
We considered models with an iron content per unit mass ranging from 0 to 80\%. Accounting for even higher amounts of metallic iron causes the silicate features to become too weak with respect
to the continuum emission that the metallic iron produces.
Silicate species with significant iron content, such as MgFeSiO$_4$, have high enough temperatures even when placed at 40 AU and require modest
or no thermal contact with metallic iron at all.

\begin{figure}[h]
   \centering
   \includegraphics[width= 8.5cm]{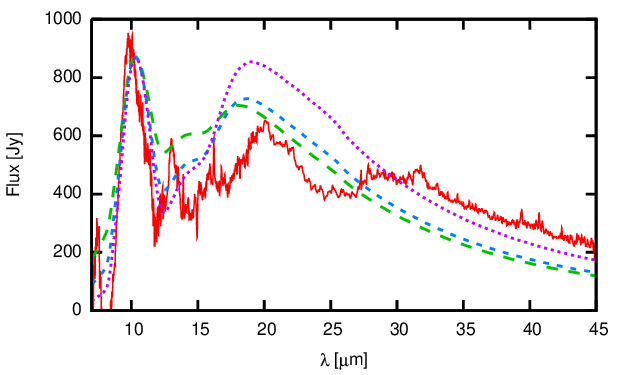}
      \caption{Models for an envelope composed of Mg$_{2}$AlSi$_{2}$O$_{7.5}$ and metallic iron ranging from 40 AU to 1800 AU. The different lines are for 30\% (dotted purple),
      50\% (short dashed blue) and 70\% (long dashed green) metallic iron content. The Al$_2$O$_3$-subtracted residual ISO spectrum is shown by the full red line.}
         \label{fig:Mg2Al}
   \end{figure}

Models were calculated for silicate envelopes ranging from 40 AU to 1800 AU (i.e. the outer limit given by the field of view of ISO SWS) for all species listed
in Table \ref{tab:optConst} separately. 
When in thermal contact with substantial amounts
of metallic iron, the silicate grains can be heated up to 400 -- 450 K at 40 AU and some of the species considered produce peak ratios that reflect the observed ratio.
As the temperatures that are reached are still fairly low, species that have relatively high values for the
$\kappa^{\rm max}_{18\,\mu{\rm m}}$/$\kappa^{\rm max}_{9.7\,\mu{\rm m}}$ are preferred.
Furthermore, metallic iron produces emission that fills the trough between the silicate peaks and matches the observed Al$_2$O$_3$-subtracted residual spectrum.
Figure \ref{fig:Mg2Al} shows the effect of varying the amount of iron in thermal contact with the silicate grains.
A higher iron content implies more emission in the silicate trough and blue-wards of the 9.7 $\mu$m peak.
For a given silicate species, the ratio between the red and the blue silicate peaks decreases with an increasing amount of metallic iron inclusions.
The dependence is stronger for smaller amounts of iron, between 0 and 50\% per mass.

When using a continuous distribution of ellipsoids \citep[CDE,][]{Bohren1998,Min2003}
as the grain model for the metallic iron particles, the amount of this species needed to heat the silicates to a given temperature is greatly reduced
if compared to models with DHS metallic iron particles.
\cite{Kemper2002} also found a large difference between the opacity of CDE metallic iron grains and that obtained using Mie theory.
We find the DHS metallic iron model to have an intermediate opacity between these two extremes and discuss the effects of using CDE metallic iron grains
in Sect. \ref{sec:silicon}.

Another feature in the Al$_2$O$_3$-subtracted residual spectrum worth noting is the broad component at 30 $\mu$m. The mass-loss rate of W\,Hya varies on
larger scales, therefore this emission bump might be thought to be a signature of an abrupt change in mass-loss rate. However, the bump seems too sharp
to be produced in such a way. Moreover, this feature is seen in many other ISO spectra of oxygen-rich AGB
stars with low mass-loss rate, suggesting a more fundamental cause. \cite{Gervais1987} studied the infrared properties of silicate glasses and concluded
that calcium-bearing silicates have a characteristic band around 30\,$\mu$m.
From the work of \cite{Mutschke1998}, we retrieved the optical data for two Ca-silicates, which do indeed show a feature at 30 $\mu$m.
Ca-bearing silicates are expected to form from gas-solid chemical reactions between Al$_2$O$_3$ grains and calcium and silicon in the gas \citep{Grossman1974}. 
Another candidate for explaining the feature is pure calcium-oxide grains.
However, this species produces a rather localized peak, whereas calcium-bearing silicates also provide continuum opacity beyond
30 $\mu$m, which better matches what is seen in the spectrum of W\,Hya.

\begin{figure}[h]
   \centering
   \includegraphics[width= 8.5cm]{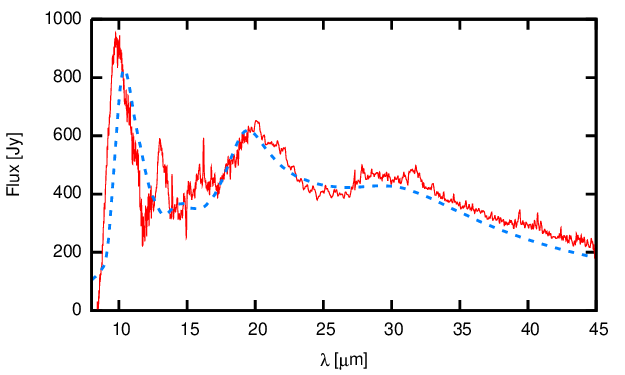}
      \caption{Fit (dashed blue) to the amorphous Al$_2$O$_3$-shell-subtracted thermal spectrum (in solid red), adopting 40 AU as the inner boundary
      of the silicate emitting envelope and a mixture of 40\% metallic iron and 60\% Ca$_{2}$Mg$_{0.5}$Al$_{2}$Si$_{1.5}$O$_{7}$ particles that are in thermal contact.}
         \label{fig:Ca2Mg0.5}
   \end{figure}

As shown in Fig. \ref{fig:Ca2Mg0.5}, the calcium-aluminum silicate
fits the Al$_2$O$_3$-subtracted residual spectral shape nicely when in thermal contact with metallic iron. Unfortunately, we were
unable to find other optical constants of calcium-bearing silicates in the literature, which potentially could provide a better fit to the 9.7\,$\mu$m silicate peak.
As calcium is indeed expected to condense just after Al$_2$O$_3$,  the existence of this dust species in the wind of W\,Hya can be expected.
The model shown in Fig. \ref{fig:Ca2Mg0.5} has a dust mass-loss rate of $3\times10^{-10}$ M$_\odot$\,year$^{-1}$, consisting of 60\% Ca$_2$Mg$_{0.5}$Al$_2$Si$_{1.5}$O$_7$ and
40\% metallic iron, per mass. Further constraints on these quantities are obtained in Sect. \ref{sec:peak70um} by fitting the broadness of the PACS 70\,$\mu$m brightness profile.

Even though the fit using Ca$_2$Mg$_{0.5}$Al$_2$Si$_{1.5}$O$_7$ is offset in the 9.7\,$\mu$m region,
this single dust species is able to qualitatively reproduce the Al$_2$O$_3$-subtracted residual spectrum very well.
Therefore we conclude that Ca$_2$Mg$_{0.5}$Al$_2$Si$_{1.5}$O$_7$ is the silicate from our set of optical constants that
best represents the composition of the silicate dust in the envelope of W\,Hya.
A better fit to the individual features might require a more complex dust envelope structure, which could be linked to the condensation of different elements at different
distances from the star, but such a more complex dust model is beyond the scope of this paper.
Having this amount of Ca$_2$Mg$_{0.5}$Al$_2$Si$_{1.5}$O$_7$ condense from the outflowing material, however, requires
a super-solar abundance of calcium and aluminum. We discuss this in Sect. \ref{sec:al+ca}.

\subsection{Fitting the peak value and broadness of the PACS 70\,$\mu$m image}
\label{sec:peak70um}

\begin{figure}[t]
   \centering
   \includegraphics[width= 8.5cm]{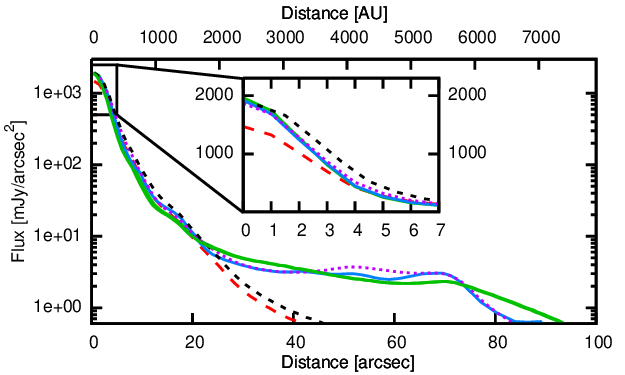}
      \caption{Fit to the radial intensity profiles observed by PACS at 70 $\mu$m.
      The blue and green solid lines represent the observed brightness distribution averaged
      over a narrow range of direction \citep{Cox2012} and over all directions, respectively. The long dashed red line represents the model
      with $\dot{M}=3\times10^{-10}$\,M$_\odot$\,year$^{-1}$, inner and outer radius of 40 AU and 1800 AU, respectively,
      and with grains composed of 40\% metallic iron and 60\% Ca$_2$Mg$_{0.5}$Al$_2$Si$_{1.5}$O$_7$.
      The short dashed black line shows the model with $\dot{M}=4\times10^{-10}$\,M$_\odot$\,year$^{-1}$, inner and outer radius of 50 AU and 1800 AU, respectively,
      and with grains composed of 50\% metallic iron and 50\% Ca$_2$Mg$_{0.5}$Al$_2$Si$_{1.5}$O$_7$.
      The dotted purple line represents the model for the recent mass-loss history obtained in Sect. \ref{sec:pacs_image}, with parameters given in Fig. \ref{fig:mass-lossHist}
      and with grains composed of 50\% metallic iron and 50\% Ca$_2$Mg$_{0.5}$Al$_2$Si$_{1.5}$O$_7$. The radial distance in AU is given along the upper x-axis for an assumed distance
      of 78 parsecs.}
         \label{fig:newPACS}
   \end{figure}

When the model consisting of an amorphous Al$_2$O$_3$ GBDS and Ca$_2$Mg$_{0.5}$Al$_2$Si$_{1.5}$O$_7$
grains in the wind that are in thermal contact with metallic iron (see Fig.~\ref{fig:newPACS}) is compared to the 
PACS 70\,$\mu$m radial brightness profile \citep{Cox2012}, the peak value of this profile
is not reproduced. The model peak flux is 30\% lower than the observed value and essentially
constrains the present-day dust mass-loss.
The broadness of the intensity profile in the first few arcseconds 
constrains the distribution of the bulk of the 70\,$\mu$m emission.

\begin{figure}[h]
   \centering
   \includegraphics[width= 8.5cm]{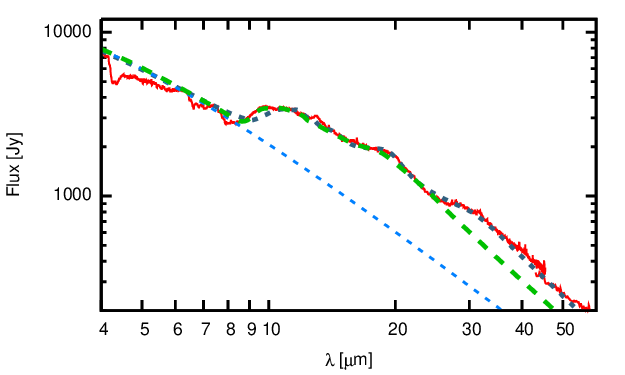}
      \caption{Overall fit of the best model (short dashed dark blue line) to the ISO SWS spectrum (solid red line). The model
       consists of the GBDS composed of amorphous Al$_2$O$_3$ grains plus an outflow (50\% Ca$_2$Mg$_{0.5}$Al$_2$Si$_{1.5}$O$_7$ and
50\% metallic iron) in which the mass-loss rate is $4\times10^{-10}$\,M$_\odot$\,year$^{-1}$ from 50 AU to 500 AU and $2\times10^{-10}$\,M$_\odot$\,year$^{-1}$ from 500 AU to 1800 AU.
The stellar input black-body spectrum is shown by a dashed light blue line. For comparison the best model from Paper I is shown by the long dashed green line.}
         \label{fig:newDustFit}
   \end{figure}

When matching the present-day dust mass loss to the peak strength, the fit to the ISO spectrum becomes poorer. 
To recover the fit, we shifted the inner radius of the silicate shell from 40 AU to 50 AU and increased the iron content
from 40\% to 50\% of the dust mass. The width of the PACS intensity profiles is matched if we decrease the
dust mass-loss rate beyond 500 AU by a factor of two, to $2\times10^{-10}$\,M$_\odot$\,year$^{-1}$, preserving 
the dust composition from the inner part. The best model thus consists of a wind composed of 50\% Ca$_2$Mg$_{0.5}$Al$_2$Si$_{1.5}$O$_7$ and
50\% metallic iron in thermal contact and mass-loss rates of $4\times10^{-10}$\,M$_\odot$\,year$^{-1}$ 
between 50 AU and 500 AU and of $2\times10^{-10}$\,M$_\odot$\,year$^{-1}$ from 500 AU to 1800 AU.
The fit to the ISO SWS spectrum of this best model for the outflow combined with the GBDS of amorphous Al$_2$O$_3$ is shown in Fig. \ref{fig:newDustFit}.
Comparing our dust model and the CO model (Paper I) yields a gas-to-dust ratio in the inner wind of W\,Hya of 325.

\subsubsection{CO envelope}

The mass-loss rate discontinuity at 500 AU is close enough to the star that it might affect the CO envelope.
The model for the CO rotational emission presented in Paper I does not take this variable mass loss into account and therefore
has to be revised.  In the CO study a fit to both the high- and low-excitation rotational lines was achieved by decreasing the
photodissociation radius of CO by a factor of 2.5, placing it at about 500 AU from W\,Hya.

\begin{figure}[t]
   \centering
   \includegraphics[width= 8.5cm]{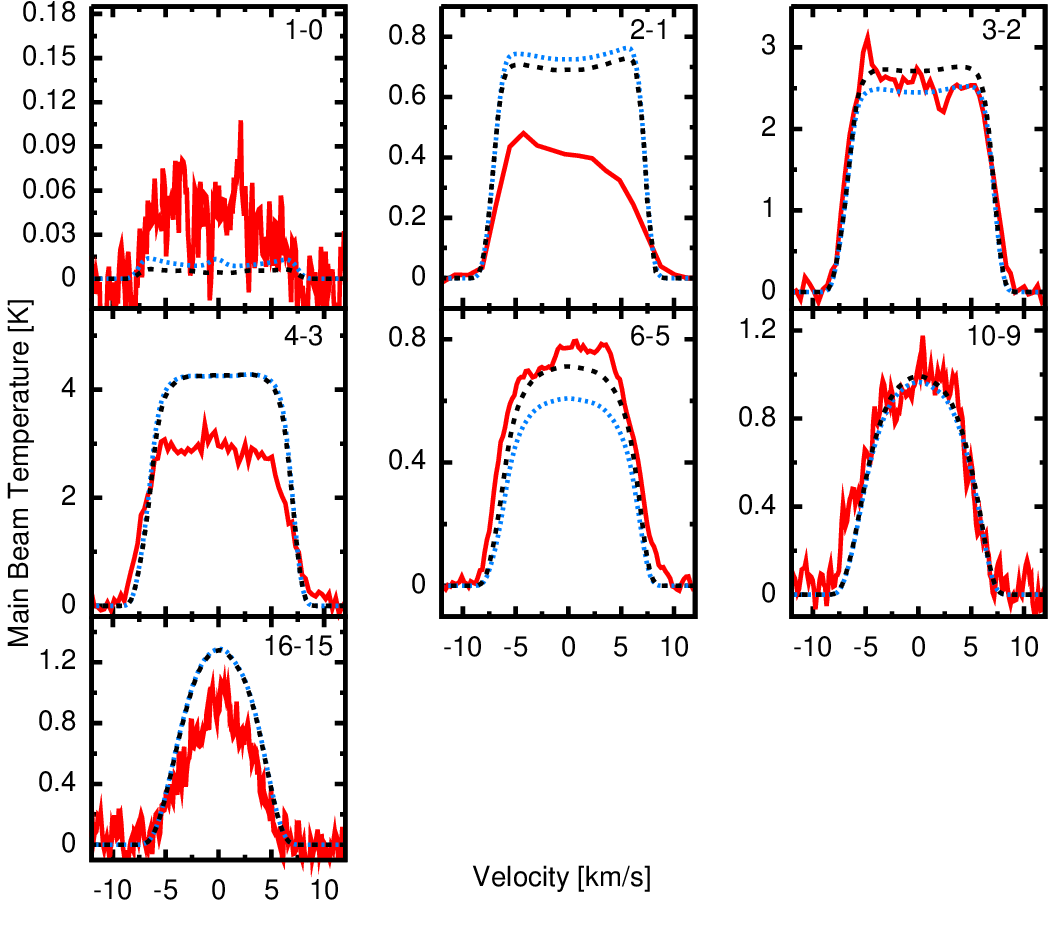}
      \caption{We compare the observed CO rotational emission lines (red solid line) to the model with mass-loss properties as described in the caption of Fig. \ref{fig:newDustFit}
      and a dissociation radius set to the canonical value obtained by \cite{Mamon1988} (shown by the blue dotted line). The model obtained in Paper I
      is shown for comparison (black dashed line).}
         \label{fig:newCO}
   \end{figure}

In Fig. \ref{fig:newCO} we compare the new model for the CO emission that adopts the new dust model and the jump
in the gas-mass-loss rate from $1.3\times10^{-7}$ to $7\times10^{-8}$\,M$_\odot$\,year$^{-1}$ at 500 AU. The new CO 
model reproduces the CO rotational lines as well as a model with CO dissociation occurring at 500 AU and thus is in
concordance with such a $\dot{M}$ discontinuity at this distance.

\subsection{Recent (dust) mass-loss history of W\,Hya}
\label{sec:pacs_image}

The mass-loss history over the past 10$^4$--10$^5$\,years is unveiled by the infrared images obtained with PACS and SPIRE and the infrared astronomical
satellite \citep[IRAS,][]{Neugebauer1984}.
The image taken by PACS at 70\,$\mu$m has the best spatial resolution and, therefore, provides the best constraints. Emission from
the extended dust envelope that seems to be produced by a higher mass-loss rate phase is also seen in the infrared ISO spectra \citep{Hawkins1990}.
This emission is only produced by cold (${\rm T}<70$\,K) dust and affects the flux measured at the relatively long wavelengths ($\lambda > 45\,\mu$m)
of the LWS spectral range; it has no impact on the dust emission in the inner wind.

\subsubsection{Infrared images}

The 70 $\mu$m PACS brightness distribution
is rather unusual for an oxygen-rich AGB star \citep{Cox2012}. These images show excess dust emission
produced between $30 ''$ and $100 ''$ that cannot be understood in terms of a constant mass-loss wind.
In Fig. \ref{fig:newPACS} we show the angle-averaged radial profile of the observed flux density at 70 $\mu$m. The PACS image shows the
envelope to have an ellipsoidal shape when projected on the sky. This causes the radial profile averaged over all directions
to have a shallower slope from around $75''$ on than radial profiles computed for a narrow range of directions.
The difference in continuum level of about 20\% between PACS and ISO at 70 $\mu$m is an effect of the significantly smaller field of view of the PACS spectrometer
compared to that of ISO LWS.

On a larger scale, 100$\,\mu$m maps obtained with IRAS show a very extended dust shell, $\approx$\,$30'$ in diameter, which requires a high
mass-loss rate to explain it \citep{Hawkins1990}.
Together, the PACS and IRAS infrared maps show that the dust mass-loss rate of W\,Hya has not been constant in the past 10$^4$ to 10$^5$ years.

\subsubsection{Model for the recent mass-loss history}

To model this extended emission, we used the inner wind dust composition and applied variations in the dust mass-loss rate.
Since the PACS image shows an elliptical dust envelope and our model assumes a spherically symmetric wind, we have measured
the radial brightness profile in a narrow range of directions (from $130^\circ$ to $170^\circ$ measured from north to east)
and used that to constrain our model. Measuring the radial profile in a different direction produces a similar curve with
the peak seen at 75$''$ in Fig. \ref{fig:newPACS} shifted to larger distances from the star. The chosen direction corresponds to the direction in which the radius of the observed
circumstellar shell is the smallest. The asymmetry seen in the PACS image could,
in principle, be due to a direction dependent expansion velocity, but that is not considered in our modelling.
The largest radius measured is larger by about 25\% than the one modelled by us. A difference of 25\% in radius translates into a difference in expansion velocity of also 25\%.

\begin{figure}[t]
   \centering
   \includegraphics[width= 8.5cm]{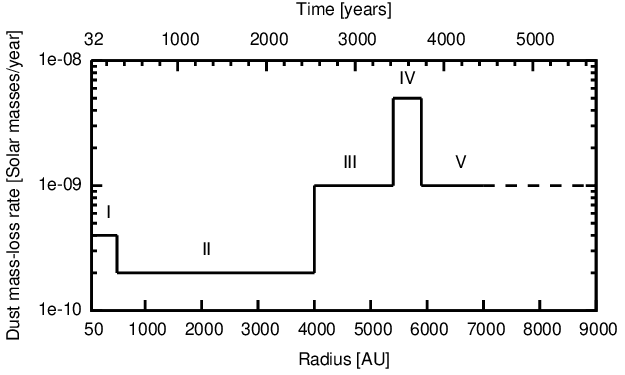}
      \caption{Schematic dust mass-loss history of W\,Hya. The mass-loss rate shown by the dashed line could not be constrained by our model.
      The timescales given along the upper x-axis is calculated assuming a constant expansion velocity of 7.5 km\ s$^{-1}$.}
         \label{fig:mass-lossHist}
   \end{figure}

The dotted purple line in Fig. \ref{fig:newPACS} shows the fit to the profile composed of a combination of the constant mass-loss rate epochs shown in Fig. \ref{fig:mass-lossHist}.
The mass-loss history and the constraints
motivating this model are given in Table \ref{tab:mass-lossHist}.
The values given assume the present-day expansion velocity of 7.5\,km\,s$^{-1}$, measured from the gas lines.
If the gas-to-dust ratio remained constant (at 325), the dust mass-loss rates of phases II, III, and IV would correspond to $6.5 \times 10^{-8}$, $3.9 \times 10^{-7}$, and
$2.0 \times 10^{-6}$ M$_\odot$\,year$^{-1}$ of gas. The exact value of the derived dust mass-loss rates depend on the model for the present-day dust mass-loss. If the composition
of the dust producing emission at $\approx\,5000$\,AU is not significantly different from that of the inner silicate envelope ($r\gtrsim50\,$AU), the derived ratio between the high and low
mass-loss rate phases should be accurate even if our inner-wind model were not. If the dust composition and/or expansion velocity changes significantly between these two components,
our results probably underestimate the mass-loss rate for the phase of stronger mass loss. That is because, first, the opacity of Ca$_2$Mg$_{0.5}$Al$_2$Si$_{1.5}$O$_7$ at 70\,$\mu$m is
higher than that of other silicates, so more mass of other silicate species at the same temperature would be required to produce the same emission at these wavelengths. And, second,
the expansion velocity tends to increase with increasing mass-loss rate \citep[see, e.g.,][]{Olofsson2002}, which means that the present value of 7.5\,km\,s$^{-1}$ would probably
be lower than the expansion velocity during a phase with higher mass-loss rate.
   
\begin{table}
\caption{Mass-loss history of W\,Hya.}             % title of Table
\label{tab:mass-lossHist}      % is used to refer this table in the text
\centering                          % used for centering table
\begin{tabular}{l @{\hspace{0.1cm}}c c l}        % centered columns (4 columns)
\hline\hline                 % inserts double horizontal lines
\rule{0pt}{2.2ex}
Phase & ${\rm \dot M_{dust}}$ & R$_{\rm in}$ & Constraint \\    % table heading 
 & [M$_\odot$\,year$^{-1}$] & [AU] &  \\    % table heading 
\hline                        % inserts single horizontal line
I & \ $4.0\times10^{-10}$ & \ \ \ \ 50 & ISO spectrum \\
II & \ $2.0\times10^{-10}$ & \ \ 500 & 70\,$\mu$m profile peak + CO model\\
III & $1.2\times10^{-9}$ & 4000 & 70\,$\mu$m profile from 20 to 60 $''$\\
IV & $6.0\times10^{-9}$ & 5400 & 70\,$\mu$m profile beyond 60 $''$\\
V & $1.2\times10^{-9}$ & 5900 & 70\,$\mu$m profile beyond 60 $''$\\
\hline                                   %inserts single line
\end{tabular}
\end{table}

\section{Discussion}
\label{sec:disc}

We have modelled the dust envelope of W\,Hya using two components: a GBDS from which the Al$_{2}$O$_{3}$ emission originates
and an outflow that contains silicates that are in thermal contact with metallic iron.  Overall, the best-fit model (see Fig. 10) provides a good fit to the infrared
emission, although the position of the silicate peak at 9.7\,$\mu$m is not well reproduced. Our model does not require a super-solar aluminum abundance that is
required if the Al$_{2}$O$_{3}$ grains that cause the emission at 12\,$\mu$m are part of the outflow.  The silicate emission originates from beyond 50 AU and best
fits the spectrum if silicate grains that contain Al and Ca are used.  This again introduces a super-solar abundance problem but we argue that this may be a
spurious problem.  We now discuss the empirically derived envelope structure in the context of wind dynamics, the gas- and solid-phase silicon budget and the
elemental abundances of Al and Ca, and the expected crystallinity of the dust grains.

\subsection{Scattering agents and wind driving in oxygen-rich AGB stars with a low mass-loss rate}
\label{sec:other_species}

The spectrum of W\,Hya is well-explained by a GBDS of amorphous Al$_2$O$_3$ that produces most of the thermal
emission seen at 12 $\mu$m. For the fit to the scattered light fractions our models can not distinguish between the two species that have been suggested,
namely silicates or Al$_2$O$_3$.
Although Al$_2$O$_3$ is able to explain the scattered light and 
thermal emission simultaneously, wind-driving models suggest that a population of pure amorphous Al$_2$O$_3$ 
grains provides insufficient opacity to initiate a wind.  Silicates, however, are much more effective in initiating the 
outflow because of the higher abundance of silicon \citep{Bladh2012}. To illustrate this: condensation of 35\% of the available silicon in a solar composition gas
into large grains of a typical silicate glass increases 
the scattering opacity by an order of magnitude relative to dust that is composed of fully condensed large amorphous 
Al$_{2}$O$_{3}$ grains alone.  If large amorphous Al$_{2}$O$_{3}$ grains are lacking in the GBDS  
the 12 $\mu$m emission is produced mainly or exclusively by small amorphous Al$_{2}$O$_{3}$ grains.

The gas mass that should be present in the GBDS to account for the amorphous Al$_2$O$_3$ emission is
$\sim$\,$10^{-5}$\,M$_\odot$.  Given the mass-loss rate of W\,Hya of $1.3 \times 10^{-7}$\,M$_\odot$\,year$^{-1}$,
this implies that only a small fraction (a few percent) of the GBDS is expelled per pulsation cycle of about a year.
The gas mass in the GBDS that is required to produce the scattered light -- assuming it is due to 35\% silicon condensing in large Mg$_2$SiO$_4$ grains 
-- is $\sim$\,$2.3\times10^{-7}$\,M$_\odot$.  This mass is more than an order of magnitude lower
than the corresponding mass for amorphous Al$_{2}$O$_{3}$, and is of about the same magnitude as
the total mass that is lost per year.
If the silicates are responsible for the scattering and condense on the amorphous Al$_2$O$_3$ seeds, condensation of this material should occur over a small radial interval of the GBDS.
This is because the composite grains must be cold not to produce silicate emission and, therefore, would also not produce any amorphous Al$_2$O$_3$ emission.
Because of this and as a result of the low abundance of aluminum, a relatively larger fraction of the GBDS volume in which warm amorphous Al$_{2}$O$_{3}$ grains exist
is needed to reproduce the observed emission.
Models of outflows driven by large Mg$_{2}$SiO$_{4}$ particles indeed show very rapid growth of these grains in the onset region of the flow \citep[see, e.g.,][]{Hofner2008}.

Alternatively, the grains responsible for the scattering could be aggregates of different oxides including species other than silicates and Al$_2$O$_3$
-- such as calcium or magnesium oxides -- but we have not explored this possibility.

\subsection{Structure of the dust envelope of W\,Hya}

\begin{figure}
\center
\begin{tikzpicture}[scale=.29]
\draw[fill=gray, color=gray] (0,0) -- (0,7) --(30,7) -- (30,-7) -- (0,-7);
\draw[fill=gray!40!white, color=gray!40!white] (0,0) -- (0,7) --(20.2,7) -- (20.2,-7) -- (0,-7);
\draw[color=gray,fill=gray!40!white, very thick] (20.21,7) arc(10.1:-10.1:40);
\draw[fill=black, color=black] (0,2.1) arc(90:-90:2.1);
\draw[fill=orange, color=orange] (0,2.0) arc(90:-90:2.0);
\draw[fill=yellow, color=yellow] (0,1.6) arc(90:-90:1.6);
\draw[fill=red, color=red] (0,1) arc(90:-90:1);
\node at (3.2,5) {STAR};
\node at (4.5,-3.5) {DUST-FREE};
\node at (5.5,3.5) {GBDS};
\node at (11,2) {SCATTERING SHELL};
\node at (15,-5) {NO SILICATE};
\node at (15,-6) {EMISSION};
\node at (26,-5) {SILICATE};
\node at (26,-6) {EMISSION};
\draw[thick] (1.5,5) -- (.3,.6);
\draw[thick] (1.3,-3.4) -- (0.55,-1.15);
\draw[thick] (3.6,3.4) -- (1.55,.9);
\draw[thick] (2.1,0) -- (5.5,2);
\end{tikzpicture}
\caption{Schematic view of the envelope of W\,Hya. The radii of the different structures are given to scale. The radius of the star and the scattering
shell are 1.8\,AU and 3.6\,AU. The inner radius of the GBDS is not constrained by our model and is assumed to be 3.1\,AU (1.7\,R$_\star$) for the small
amorphous Al$_2$O$_3$ grains. The inner radius of the silicate envelope shown is the lower limit of 40\,AU given by \cite{Zhao-Geisler2011}.}
\label{fig:schem}
\end{figure}
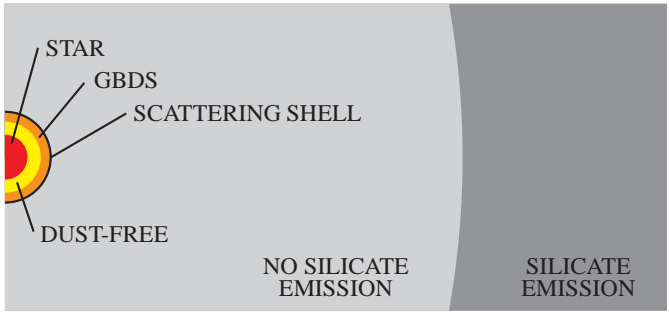

A schematic view of the model for the envelope of W\,Hya is shown in Fig. \ref{fig:schem}. Most of the amorphous Al$_2$O$_3$ emission can be accounted for by
grains in the GBDS. The thickness of this GBDS is difficult to assess as its inner radius depends on the poorly constrained near-infrared optical constants of amorphous
Al$_2$O$_3$. At about 2\,R$_\star$ grains of 0.3\,$\mu$m are present to account for the scattered light fractions observed. In our models, these grains
can be either amorphous Al$_2$O$_3$ or iron-free silicates. If these scatterers are silicates, it is likely that
the amorphous Al$_2$O$_3$ grains serve as seeds for silicate condensation \citep{Kozasa1997}.
If silicates exist so close to the star, they have to be translucent in the near-infrared and remain cold such as not to produce significant emission.
Combined with the broadness of the brightness peak in the PACS 70\,$\mu$m image,
this implies that the bulk of silicate emission at 9.7\,$\mu$m has to come from an envelope with an inner radius of 50\,AU. The gas temperature
at this distance is $\sim$350\,K \citep{Khouri2014a}.

 If the silicates are the scatterers and are present near the star without producing significant emission, there would have to be a
physical process that increases their temperature beyond 40\,AU.
Such a process could be a change in their lattice structure, which increases the absorption in the near-infrared, or the establishing of contact with a more opaque dust
species, for instance small iron grains -- creating a silicate with iron inclusions.
Alternatively, if the scattering is caused by Al$_2$O$_3$, or grains composed of a combination of other oxides, the inner radius of the silicate emission envelope
could be set by silicate condensation. It is not clear, however, why any of the processes mentioned would occur at such large distances.
This interpretation is based on the assumption of a steady-state wind model for the inner regions, $r < 1000$\,AU. The observed radial dependency of the dust-envelope
parameters of W\,Hya could potentially also be the result of variable conditions at the wind acceleration region, but this explanation seems less likely (see Sect. \ref{sec:expan}).

\subsection{Bringing the dust and gas models together: elemental abundances in the inner wind.}

The abundances of the elemental species that are required to reproduce the thermal emission from
the wind of W\,Hya are 4.8 times the solar-composition value for calcium, 0.06 for magnesium, 3.5 for aluminum, 0.24 for silicon, 0.07 for oxygen, and 1.15 for iron.
If the opacity of the metallic iron grains is computed using a CDE distribution of sizes,
a factor of three less iron is needed.  In what follows, we discuss the abundances of the dust 
species in more detail and offer potential solutions to the apparent super-solar abundances of calcium and aluminum.

\subsubsection{Silicon abundance}
\label{sec:silicon}

\cite{Khouri2014a} modelled the gas-phase SiO emission and found that between 15 and 50\% of the silicon atoms are expected to be depleted from the gas-phase.
When considered together with our dust model this
accounts for all the silicon expected for a solar composition wind.

However, since the calcium-bearing silicate that we adopted (Ca$_2$Mg$_{0.5}$Al$_2$Si$_{1.5}$O$_7$)
does not provide a perfect fit to the 9.7\,$\mu$m silicate peak, the constraints on the solid-phase silicon abundance
are not stringent. To assess this problem, we have compared models of other silicates that provide a better fit to the 9.7\,$\mu$m peak silicate feature to the observed spectrum.
These dust species provide a poorer fit to the overall spectrum, however, as they reproduce neither the
18\,$\mu$m silicate peak nor the 30 $\mu$m feature. We have chosen to fit the flux ratio between the 9.7\,$\mu$m silicate peak and the trough at 11.9 and 14.5 $\mu$m.
The trough flux is sensitive to the amount of iron in thermal contact with the silicates, which needs to be 50\% per mass, typically, in our models.
The best fits are always achieved with in between 30\% and 40\% of the silicon available from the gas phase.
Therefore, the amount of silicon needed in the dust agrees with the depletion fraction reported by \cite{Khouri2014a} independent of the exact
composition of the silicate.
This is in virtue of the strength of the 9.7\,$\mu$m peak correlating well with the SiO$_2$ content of the given silicate species \citep[e.g.][]{Koike1987}.

\subsubsection{Aluminum and calcium abundances}
\label{sec:al+ca}

In our model, aluminum is condensed in amorphous Al$_2$O$_3$ grains in the GBDS and
also in silicates present in the wind. We do not require amorphous Al$_2$O$_3$ to be present in the
wind but a significant amount in terms of aluminum abundance could be driven out without
producing a strong spectral signature. We considered only the outflowing aluminum-bearing grains in the aluminum budget,
as the grains that form in the GBDS form from a gravitationally bound gas mass that is not independently constrained by our models.
Nonetheless, our dust model requires 3.5 times more aluminum than what is available from the gas phase.
Unfortunately, we do not have aluminum-free calcium-bearing silicates in our dataset of optical constants, so we cannot directly probe the impact of the
aluminum content on the optical properties of Ca$_2$Mg$_{0.5}$Al$_2$Si$_{1.5}$O$_7$. \cite{Gervais1987} studied the influence of
aluminum content on silicate emission features and found it to have a minor impact on the strongest ressonances. This suggests that
the over-abundance of aluminum found by us may be spurious and that an equally good fit to the Al$_2$O$_3$-subtracted residual spectrum could be achieved
with a silicate with a lower aluminum content.

The required super-solar abundance of calcium is somewhat more intriguing, as this element
seems to have a significant impact on the strengths and shapes of the silicate features. Particularly the 30 $\mu$m feature is linked to the presence of this element.
Like in the case of aluminum, however, we do not have a library of optical constants for silicate species with different amounts of calcium content,
which prevents us from assessing its relation to the 30 $\mu$m feature.

We identify three potential solutions to the problem of the super-solar abundance of aluminum and calcium in the dust found by us.
The first option is that silicates that contain smaller amounts of calcium and aluminum are responsible for the thermal emission.
\cite{Speck2011} presented measurements of silicates with a range of compositions and calcium and aluminum content. One of the species studied by
them, named basalt in their work, has calcium-to-silicon and aluminum-to-silicon ratios very similar to those expected for 20\% silicon condensation
and full calcium and aluminum condensation. The absorbance spectrum of this species shown in Fig. 4 of their paper peaks at 10 $\mu$m, which is
closer to the observed peak position for W\,Hya than that of Ca$_2$Mg$_{0.5}$Al$_2$Si$_{1.5}$O$_7$.  The 18 $\mu$m
peak of this species is also in accordance with the observed spectrum of W\,Hya. Unfortunately for our purposes, the authors focused on analysing the 10 $\mu$m and 18 $\mu$m
region and we have no information on the spectra of this species at 30 $\mu$m.
Measurements of optical data for silicates with these characteristics are needed to determine whether the spectrum of W\,Hya can be reproduced by such a species.

A second possibility to explain the apparent super-solar abundance of Al and Ca is that some of the features fitted by Ca$_2$Mg$_{0.5}$Al$_2$Si$_{1.5}$O$_7$,
such as those at 20\,$\mu$m and 30\,$\mu$m, are partially produced
in the GBDS, similarly to Al$_2$O$_3$. In that case the abundance problem would not exist. This can be the case if other oxides are able to condense in the GBDS.
Calcium is indeed expected to be one of the first elements to react with Al$_2$O$_3$ from condensation calculations \citep{Grossman1974}.
If calcium reacts with and is incorporated into the Al$_2$O$_3$ grains before the onset of the wind, other features in the spectrum could also be produced
in the extended atmosphere. Furthermore, the addition of calcium to the dust grains in the GBDS
could help driving the wind. However, the increase in the dust-to-gas ratio due to calcium condensation would only be of about a factor of two.

Finally, calcium and aluminum could be overabundant in the wind if these atoms are included in the dust grains before the wind
is fully driven and if only a fraction of the gas is driven out by these grains. Although this might be the case, this scenario seems unlikely as the dust grains should be well
coupled to the gas at the densities expected in the region where the flow is initiated.

\subsection{Lattice structure of the grains in the GBDS}

The grains in the GBDS must form from high-temperature gas and are heated by the stellar
radiation field to temperatures that may differ from that of the gas. Being subjected to high enough temperatures with respect to the glass temperature of the given dust species will cause the dust grains
to become crystalline. Whether the lattice structures of the dust grains are amorphous or crystalline is directly connected to the features
produced by them in the infrared spectrum. In particular, our model requires that amorphous Al$_2$O$_3$ grains exist close to the star
($\sim$\,2\,R$_\star$) to reproduce the ISO spectrum.
As we obtain the grain temperature as a function of radius, we can speculate what the lattice structure
of Al$_2$O$_3$ and silicate grains in the GBDS would be and test if that is consistent with our assumptions.

\subsubsection{Al$_2$O$_3$ grains}
\label{sec:al2o3_lattice}

The physical conditions in AGB outflows favour the condensation of amorphous Al$_2$O$_3$ grains \citep[e.g.][]{DellAgli2014}.
In our calculations the temperature of the grains in the GBDS is found to be between 1500 K and 1600 K for large grains and between 1200 K and 1300 K for small grains.
\cite{Begemann1997} found that at 1300 K Al$_2$O$_3$ grains are crystalline, while \cite{Levin1998} reported that the crystalline component
only dominates at higher temperatures ($\sim$\,1450\,K). The optical data at short wavelengths for Al$_2$O$_3$
are quite uncertain, which translates into an uncertainty in the calculated temperature of the amorphous Al$_2$O$_3$ grains.
For instance, a lower opacity in the near-infrared would cause the grains to be cooler.
If we consider the size and location of the amorphous Al$_2$O$_3$ shell and maintain the amounts of grains, the emission produced is proportional to the Planck function,
since the emitting medium is optically thin. Therefore, a lower temperature of 1400 or 1300 K for the large grains would correspond to 10 or 20\% less emission, respectively, and the
amorphous Al$_2$O$_3$ from the shell could still dominate the emission at 12\,$\mu$m.

It is also possible, and quite likely, that crystalline and  amorphous Al$_2$O$_3$ co-exist in the envelope of oxygen-rich AGB stars. In fact, grains consisting of pure crystalline
Al$_2$O$_3$ \citep{Zeidler2013} and those consisting of a crystalline Al$_2$O$_3$ core and a silicate mantle \citep{Kozasa1997a}
have been proposed as the carriers of the unidentified 13\,$\mu$m feature. We tried to fit the 13\,$\mu$m feature of W\,Hya with pure crystalline Al$_2$O$_3$ grains in the GBDS using
the optical constant measured by \cite{Zeidler2013}.
We can only obtain a good fit  by considering spherical (rather than DHS) particles and the optical constants for crystalline
Al$_2$O$_3$ grains at 973\,K. Our models show that the mass abundance of
crystalline Al$_2$O$_3$ grains compared to the amorphous ones would be very small, around 5\% by mass. We note that
\cite{Takigawa2014} found a high fraction of crystalline pre-solar Al$_2$O$_3$ grains. The authors measured five out of nine studied
pre-solar Al$_2$O$_3$ grains, which were probably produced in the ejecta of a low-mass AGB star. The derived degree of crystallinity
is considerably higher than that inferred from the model presented in this work.

\subsubsection{Silicate grains}

We can speculate on what would be the lattice structure of silicate solids if they condensed on a warm Al$_2$O$_3$ core in the GBDS.
If a silicate solid is heated to a temperature above its glass temperature ($\sim$\,1000\,K), it will crystallize.
Large pure Mg$_2$SiO$_4$ grains have a temperature of 700\,K at 2.0\,R$_\star$; a population of small pure amorphous Al$_2$O$_3$ grains at the same distance is
heated to about 1300\,K. The temperature of a composite grain will obviously depend on the relative fraction of each species. 
If a low mass of Mg$_2$SiO$_4$ condenses onto the surface of an amorphous Al$_2$O$_3$ core
the Mg$_2$SiO$_4$ solid is expected to be crystalline because the temperature of the composite would rise above the glass temperature for silicates.
However, if 35\% silicon condensation and full aluminum condensation is assumed, the silicate grain mass would be approximately 11.5 times
that of amorphous Al$_2$O$_3$ and the composite grains would probably remain cold and undetectable.
A rough calculation suggests that when the mass of Mg$_2$SiO$_4$ and amorphous Al$_2$O$_3$ are comparable, the temperature of the composite
grain would be about 1000\,K at 2.0\,R$_\star$, that is close to the glass temperature for silicates. This means that only roughly the first 10\%
of silicate material to condense onto an amorphous Al$_2$O$_3$ grain in the GBDS would yield a crystalline solid, while continued silicate condensation would make
the grain amorphous.

\subsection{Expansion velocity profile of the wind of W\,Hya}
\label{sec:expan}

In our best-fit model, the silicate emission is produced in an envelope with inner radius of 50 AU. In order for these grains to be visible, large amounts of metallic iron must be
placed in thermal contact with the silicate grains.
We can connect this finding with the gas model from Paper I. In that work, the wind of W\,Hya was found to slowly accelerate up to about 5.5 km/s and then to quickly accelerate
to the final expansion velocity of 7.5 km/s. Such an additional acceleration requires extra opacity to add momentum to the (supersonic) flow. In a dust-driven outflow,
changes in the dust properties could in principle provide such additional momentum.
We speculate that the late condensation of metallic iron on the silicate grains might provide the extra opacity needed for an increase in the wind speed to its final value.

This interpretation is based on a steady-state picture for the inner wind ($r\lesssim1000$\ AU). An alternative explanation is that the conditions at the
wind-acceleration region have changed about 30 years ago, causing the dust composition to change from silicate-rich to silicate-poor and the wind expansion velocity to decrease.
However, we do not see signs of significant mass-loss rate variations in the acceleration zone, as one might expect if the local conditions had significantly changed.
In addition, the expansion velocity measured using different diagnostics has been shown to increase steadily outwards \citep[see, e.g.,][]{Muller2008,Szymczak1998}.
Moreover, if the material at $r>50$\,AU and now expanding at 7.5\,km\,s$^{-1}$ had been ejected with
a higher velocity from the start than material inwards, one would expect to see a gap (of about 15\,AU) between these two wind components.
There is no support for such a gap in the data at hand. Although we are not able to rule out this alternative explanation completely, we consider it unlikely.

\subsection{Recent mass-loss history of W\,Hya}

We find that W\,Hya has undergone abrupt changes of more than an order of magnitude in its dust mass-loss rate in the past few thousand years.
\cite{Hawkins1990} reported observations carried out with IRAS that reveal a dust shell of $1\times10^{-4}$\,M$_\odot$ extending to about $15\,'$.
If the same expansion velocity is considered, it would take the dust grains forty-five thousand years to reach such distances from the star and, therefore, the average dust mass-loss
rate in that period would be $2\times10^{-9}$\,M$_\odot$\,year$^{-1}$. This value is compatible with the mass-loss rate derived by us for the phases with high mass-loss rates (phases
II, IV and V, given in Table \ref{tab:mass-lossHist}),
rather than to the more recent period of low mass-loss rate (phases
I and II, given in Table \ref{tab:mass-lossHist}).
Therefore, if the mass loss fluctuates between a low and high mode, the high mode must be the norm. W\,Hya might
have recently undergone a change in mass-loss rate that had not occurred in the past fifty thousand years.

Although thin shells produced by similar mass-loss rate discontinuities are found somewhat frequently around carbon-rich AGB stars \citep[e.g.][]{Olofsson1990,Olofsson1996},
the narrow mass-loss rate peak seen between 5400 and 5900 AU in the PACS images of W\,Hya is unusual for oxygen-rich AGB stars \citep{Cox2012}.
The shells around carbon-stars are thought to form as a result of a strong increase in mass-loss rate during a thermal-pulse \citep{Schroeder1998,Mattsson2007}. It is not known
why these detached shells are not observed around oxygen-rich AGB stars, and W\,Hya is unique in that sense.
The zero-age-main-sequence
mass of W\,Hya was determined to be $1.5\pm0.2$\,M$_\odot$ \citep{Khouri2014a}, which places this object around the limit for becoming a carbon-rich star \citep{Wallerstein1998}.
However, models for the formation of detached shells require the interaction of a faster wind with a previously slower wind \citep[e.g.][]{Schoier2005a}, and that is not expected
to be directly connected to stellar mass, but rather to changes in the wind properties during the thermal pulse.
Hence, it is currently not clear whether the peak in the mass-loss rate of W\,Hya is due to a thermal pulse or whether the star can be seen as an oxygen-rich analogue to carbon
stars with detached shells.

\section{Summary}
\label{sec:summary}

We presented a model for the dust envelope of W\,Hya that simultaneously considers different observables of the dust envelope of this star:
the scattered light fractions \citep{Norris2012}, the ISO spectrum \citep{Justtanont2004,Sloan2003}, the inner radius of the silicate emission shell
\citep{Zhao-Geisler2011}, the elemental abundances in the wind \citep[Paper I, ][]{Khouri2014a}, and the PACS 70\,$\mu$m image \citep{Cox2012}.

Our model consists of a gravitationally bound dust shell combined with an outflow.
An MRN distribution of amorphous Al$_2$O$_3$ particles in the GBDS fits both the infrared excess emission
and the scattered light fractions.
However, large translucent silicates that show no spectral signature at $9.7\,\mu$m, but that are able to reproduced the scattered light fractions,
can exist in the GBDS and wind, implying that we are unable to determine the nature of the scattering agents.
Silicate grains are preferred, however, based on wind-driving models because of the low abundance of aluminum \citep{Hofner2008, Bladh2012}.
Independent of the nature of the grains that drive the outflow, spectral emission from the silicate grains arises from
distances beyond 50\,AU from the star. This emission may arise when iron inclusions are incorporated into the silicates heating up the grains.
The silicon content that is in the dust is found to be consistent with the amount of silicon atoms that disappears from the gas phase,
as extracted from SiO line emission \citep{Khouri2014a}.
We find evidence that calcium is a constituent of the dust in the wind of W\,Hya.
It is difficult, however, to quantify the calcium content of the dust because we only have optical constants available for three calcium-bearing species.
Although our best fit seems to suggest a super-solar Ca abundance in the outflow,
optical data over a broad wavelength range of silicates with lower calcium content might result in fits with a solar calcium abundance.
Emission from calcium-bearing dust species might also originate from the GBDS, as long as there is no strong contribution in the MIDI wavelength range.

The mass-loss rate of W\,Hya seems to have been at the present level for about 300 years. At times before this, we see a phase when
the mass-loss was lower by a factor of roughly two, which is traced by both the dust envelope and the CO gas emission. The mass-loss rate remained at this level
for about 2500 years and produced dust now located in the envelope between 500 and 4000 AU. This phase of lower mass loss was preceded by 
one where W\,Hya lost mass at a rate at least three times higher than at present, with a localised peak in which the rate was
even twenty times higher and which ended about 3400 years ago and lasted for about 300 years.

 \begin{acknowledgements}
 The authors thank the anonymous referee for the careful reading of the paper and the
 suggestions that helped improve the quality and clarity of the text. 
 PACS has been developed by a consortium of institutes
led by MPE (Germany) and including UVIE (Austria); KUL, CSL,
IMEC (Belgium); CEA, OAMP (France); MPIA (Germany); IFSI, OAP/AOT,
OAA/CAISMI, LENS, SISSA (Italy); IAC (Spain). This development has been
supported by the funding agencies BMVIT (Austria), ESA-PRODEX (Belgium),
CEA/CNES (France), DLR (Germany), ASI (Italy), and CICYT/MCYT (Spain).
T.Kh. gratefully acknowledges the support from NWO grant 614.000.903.
N.L.J.C. acknowledges support from the Belgian Federal Science Policy Office via the PRODEX Programme of ESA.
 \end{acknowledgements}

\bibliographystyle{aa}
\bibliography{../../bibliography}

\begin{thebibliography}{67}
\expandafter\ifx\csname natexlab\endcsname\relax\def\natexlab#1{#1}\fi

\bibitem[{{Asplund} {et~al.}(2009){Asplund}, {Grevesse}, {Sauval}, \&
  {Scott}}]{Asplund2009}
{Asplund}, M., {Grevesse}, N., {Sauval}, A.~J., \& {Scott}, P. 2009, \araa, 47,
  481

\bibitem[{{Begemann} {et~al.}(1997){Begemann}, {Dorschner}, {Henning},
  {Mutschke}, {Guertler}, {Koempe}, \& {Nass}}]{Begemann1997}
{Begemann}, B., {Dorschner}, J., {Henning}, T., {et~al.} 1997, \apj, 476, 199

\bibitem[{{Bladh} \& {H{\"o}fner}(2012)}]{Bladh2012}
{Bladh}, S. \& {H{\"o}fner}, S. 2012, \aap, 546, A76

\bibitem[{{Bladh} {et~al.}(2013){Bladh}, {H{\"o}fner}, {Nowotny}, {Aringer}, \&
  {Eriksson}}]{Bladh2013}
{Bladh}, S., {H{\"o}fner}, S., {Nowotny}, W., {Aringer}, B., \& {Eriksson}, K.
  2013, \aap, 553, A20

\bibitem[{{Bohren} \& {Huffman}(1998)}]{Bohren1998}
{Bohren}, C.~F. \& {Huffman}, D.~R. 1998, {Absorption and Scattering of Light
  by Small Particles}

\bibitem[{{Cami}(2002)}]{Cami2002}
{Cami}, J. 2002, PhD thesis, University of Amsterdam

\bibitem[{{Clegg} {et~al.}(1996){Clegg}, {Ade}, {Armand}, {Baluteau}, {Barlow},
  {Buckley}, {Berges}, {Burgdorf}, {Caux}, {Ceccarelli}, {Cerulli}, {Church},
  {Cotin}, {Cox}, {Cruvellier}, {Culhane}, {Davis}, {di Giorgio}, {Diplock},
  {Drummond}, {Emery}, {Ewart}, {Fischer}, {Furniss}, {Glencross},
  {Greenhouse}, {Griffin}, {Gry}, {Harwood}, {Hazell}, {Joubert}, {King},
  {Lim}, {Liseau}, {Long}, {Lorenzetti}, {Molinari}, {Murray}, {Naylor},
  {Nisini}, {Norman}, {Omont}, {Orfei}, {Patrick}, {Pequignot}, {Pouliquen},
  {Price}, {Nguyen-Q-Rieu}, {Rogers}, {Robinson}, {Saisse}, {Saraceno},
  {Serra}, {Sidher}, {Smith}, {Smith}, {Spinoglio}, {Swinyard}, {Texier},
  {Towlson}, {Trams}, {Unger}, \& {White}}]{Clegg1996}
{Clegg}, P.~E., {Ade}, P.~A.~R., {Armand}, C., {et~al.} 1996, \aap, 315, L38

\bibitem[{{Cox} {et~al.}(2012){Cox}, {Kerschbaum}, {van Marle}, {Decin},
  {Ladjal}, {Mayer}, {Groenewegen}, {van Eck}, {Royer}, {Ottensamer}, {Ueta},
  {Jorissen}, {Mecina}, {Meliani}, {Luntzer}, {Blommaert}, {Posch},
  {Vandenbussche}, \& {Waelkens}}]{Cox2012}
{Cox}, N.~L.~J., {Kerschbaum}, F., {van Marle}, A.-J., {et~al.} 2012, \aap,
  537, A35

\bibitem[{{de Graauw} {et~al.}(1996){de Graauw}, {Haser}, {Beintema},
  {Roelfsema}, {van Agthoven}, {Barl}, {Bauer}, {Bekenkamp}, {Boonstra},
  {Boxhoorn}, {Cote}, {de Groene}, {van Dijkhuizen}, {Drapatz}, {Evers},
  {Feuchtgruber}, {Frericks}, {Genzel}, {Haerendel}, {Heras}, {van der Hucht},
  {van der Hulst}, {Huygen}, {Jacobs}, {Jakob}, {Kamperman}, {Katterloher},
  {Kester}, {Kunze}, {Kussendrager}, {Lahuis}, {Lamers}, {Leech}, {van der
  Lei}, {van der Linden}, {Luinge}, {Lutz}, {Melzner}, {Morris}, {van Nguyen},
  {Ploeger}, {Price}, {Salama}, {Schaeidt}, {Sijm}, {Smoorenburg}, {Spakman},
  {Spoon}, {Steinmayer}, {Stoecker}, {Valentijn}, {Vandenbussche}, {Visser},
  {Waelkens}, {Waters}, {Wensink}, {Wesselius}, {Wiezorrek}, {Wieprecht},
  {Wijnbergen}, {Wildeman}, \& {Young}}]{deGraauw1996}
{de Graauw}, T., {Haser}, L.~N., {Beintema}, D.~A., {et~al.} 1996, \aap, 315,
  L49

\bibitem[{{Dell'Agli} {et~al.}(2014){Dell'Agli}, {Garc{\'{\i}}a-Hern{\'a}ndez},
  {Rossi}, {Ventura}, {Di Criscienzo}, \& {Schneider}}]{DellAgli2014}
{Dell'Agli}, F., {Garc{\'{\i}}a-Hern{\'a}ndez}, D.~A., {Rossi}, C., {et~al.}
  2014, ArXiv e-prints

\bibitem[{{Dorschner} {et~al.}(1995){Dorschner}, {Begemann}, {Henning},
  {Jaeger}, \& {Mutschke}}]{Dorschner1995}
{Dorschner}, J., {Begemann}, B., {Henning}, T., {Jaeger}, C., \& {Mutschke}, H.
  1995, \aap, 300, 503

\bibitem[{{Fabian} {et~al.}(2001){Fabian}, {Posch}, {Mutschke}, {Kerschbaum},
  \& {Dorschner}}]{Fabian2001}
{Fabian}, D., {Posch}, T., {Mutschke}, H., {Kerschbaum}, F., \& {Dorschner}, J.
  2001, \aap, 373, 1125

\bibitem[{{Gail} \& {Sedlmayr}(1999)}]{Gail1999}
{Gail}, H.-P. \& {Sedlmayr}, E. 1999, \aap, 347, 594

\bibitem[{{Gail} {et~al.}(2013){Gail}, {Wetzel}, {Pucci}, \&
  {Tamanai}}]{Gail2013}
{Gail}, H.-P., {Wetzel}, S., {Pucci}, A., \& {Tamanai}, A. 2013, \aap, 555,
  A119

\bibitem[{{Gervais} {et~al.}(1987){Gervais}, {Blin}, {Massiot}, {Coutures},
  {Chopinet}, \& {Naudin}}]{Gervais1987}
{Gervais}, F., {Blin}, A., {Massiot}, D., {et~al.} 1987, Journal of Non
  Crystalline Solids, 89, 384

\bibitem[{{Griffin} {et~al.}(2010){Griffin}, {Abergel}, {Abreu}, {Ade},
  {Andr{\'e}}, {Augueres}, {Babbedge}, {Bae}, {Baillie}, {Baluteau}, {Barlow},
  {Bendo}, {Benielli}, {Bock}, {Bonhomme}, {Brisbin}, {Brockley-Blatt},
  {Caldwell}, {Cara}, {Castro-Rodriguez}, {Cerulli}, {Chanial}, {Chen},
  {Clark}, {Clements}, {Clerc}, {Coker}, {Communal}, {Conversi}, {Cox},
  {Crumb}, {Cunningham}, {Daly}, {Davis}, {de Antoni}, {Delderfield}, {Devin},
  {di Giorgio}, {Didschuns}, {Dohlen}, {Donati}, {Dowell}, {Dowell}, {Duband},
  {Dumaye}, {Emery}, {Ferlet}, {Ferrand}, {Fontignie}, {Fox}, {Franceschini},
  {Frerking}, {Fulton}, {Garcia}, {Gastaud}, {Gear}, {Glenn}, {Goizel},
  {Griffin}, {Grundy}, {Guest}, {Guillemet}, {Hargrave}, {Harwit}, {Hastings},
  {Hatziminaoglou}, {Herman}, {Hinde}, {Hristov}, {Huang}, {Imhof}, {Isaak},
  {Israelsson}, {Ivison}, {Jennings}, {Kiernan}, {King}, {Lange}, {Latter},
  {Laurent}, {Laurent}, {Leeks}, {Lellouch}, {Levenson}, {Li}, {Li},
  {Lilienthal}, {Lim}, {Liu}, {Lu}, {Madden}, {Mainetti}, {Marliani}, {McKay},
  {Mercier}, {Molinari}, {Morris}, {Moseley}, {Mulder}, {Mur}, {Naylor},
  {Nguyen}, {O'Halloran}, {Oliver}, {Olofsson}, {Olofsson}, {Orfei}, {Page},
  {Pain}, {Panuzzo}, {Papageorgiou}, {Parks}, {Parr-Burman}, {Pearce},
  {Pearson}, {P{\'e}rez-Fournon}, {Pinsard}, {Pisano}, {Podosek}, {Pohlen},
  {Polehampton}, {Pouliquen}, {Rigopoulou}, {Rizzo}, {Roseboom}, {Roussel},
  {Rowan-Robinson}, {Rownd}, {Saraceno}, {Sauvage}, {Savage}, {Savini},
  {Sawyer}, {Scharmberg}, {Schmitt}, {Schneider}, {Schulz}, {Schwartz},
  {Shafer}, {Shupe}, {Sibthorpe}, {Sidher}, {Smith}, {Smith}, {Smith},
  {Spencer}, {Stobie}, {Sudiwala}, {Sukhatme}, {Surace}, {Stevens}, {Swinyard},
  {Trichas}, {Tourette}, {Triou}, {Tseng}, {Tucker}, {Turner}, {Vaccari},
  {Valtchanov}, {Vigroux}, {Virique}, {Voellmer}, {Walker}, {Ward}, {Waskett},
  {Weilert}, {Wesson}, {White}, {Whitehouse}, {Wilson}, {Winter}, {Woodcraft},
  {Wright}, {Xu}, {Zavagno}, {Zemcov}, {Zhang}, \& {Zonca}}]{Griffin2010}
{Griffin}, M.~J., {Abergel}, A., {Abreu}, A., {et~al.} 2010, \aap, 518, L3

\bibitem[{{Grossman} \& {Larimer}(1974)}]{Grossman1974}
{Grossman}, L. \& {Larimer}, J.~W. 1974, Reviews of Geophysics and Space
  Physics, 12, 71

\bibitem[{{Habing}(1996)}]{Habing1996}
{Habing}, H.~J. 1996, \aapr, 7, 97

\bibitem[{{Habing} \& {Olofsson}(2003)}]{Habing2003}
{Habing}, H.~J. \& {Olofsson}, H., eds. 2003, {Asymptotic Giant Branch Stars}

\bibitem[{{Hawkins}(1990)}]{Hawkins1990}
{Hawkins}, G.~W. 1990, \aap, 229, L5

\bibitem[{{Henning} {et~al.}(1995){Henning}, {Begemann}, {Mutschke}, \&
  {Dorschner}}]{Henning1995}
{Henning}, T., {Begemann}, B., {Mutschke}, H., \& {Dorschner}, J. 1995, \aaps,
  112, 143

\bibitem[{{Heras} \& {Hony}(2005)}]{Heras2005}
{Heras}, A.~M. \& {Hony}, S. 2005, \aap, 439, 171

\bibitem[{{H{\"o}fner}(2008)}]{Hofner2008}
{H{\"o}fner}, S. 2008, \aap, 491, L1

\bibitem[{{Ireland} {et~al.}(2004){Ireland}, {Tuthill}, {Bedding}, {Robertson},
  \& {Jacob}}]{Ireland2004}
{Ireland}, M.~J., {Tuthill}, P.~G., {Bedding}, T.~R., {Robertson}, J.~G., \&
  {Jacob}, A.~P. 2004, \mnras, 350, 365

\bibitem[{{J{\"a}ger} {et~al.}(2003){J{\"a}ger}, {Dorschner}, {Mutschke},
  {Posch}, \& {Henning}}]{Jager2003}
{J{\"a}ger}, C., {Dorschner}, J., {Mutschke}, H., {Posch}, T., \& {Henning}, T.
  2003, \aap, 408, 193

\bibitem[{{J{\"a}ger} {et~al.}(1994){J{\"a}ger}, {Mutschke}, {Begemann},
  {Dorschner}, \& {Henning}}]{Jager1994}
{J{\"a}ger}, C., {Mutschke}, H., {Begemann}, B., {Dorschner}, J., \& {Henning},
  T. 1994, \aap, 292, 641

\bibitem[{{Justtanont} {et~al.}(2004){Justtanont}, {de Jong}, {Tielens},
  {Feuchtgruber}, \& {Waters}}]{Justtanont2004}
{Justtanont}, K., {de Jong}, T., {Tielens}, A.~G.~G.~M., {Feuchtgruber}, H., \&
  {Waters}, L.~B.~F.~M. 2004, \aap, 417, 625

\bibitem[{{Justtanont} \& {Tielens}(1992)}]{Justtanont1992}
{Justtanont}, K. \& {Tielens}, A.~G.~G.~M. 1992, \apj, 389, 400

\bibitem[{{Karovicova} {et~al.}(2013){Karovicova}, {Wittkowski}, {Ohnaka},
  {Boboltz}, {Fossat}, \& {Scholz}}]{Karovicova2013}
{Karovicova}, I., {Wittkowski}, M., {Ohnaka}, K., {et~al.} 2013, \aap, 560, A75

\bibitem[{{Kemper} {et~al.}(2002){Kemper}, {de Koter}, {Waters}, {Bouwman}, \&
  {Tielens}}]{Kemper2002}
{Kemper}, F., {de Koter}, A., {Waters}, L.~B.~F.~M., {Bouwman}, J., \&
  {Tielens}, A.~G.~G.~M. 2002, \aap, 384, 585

\bibitem[{{Kessler} {et~al.}(1996){Kessler}, {Steinz}, {Anderegg}, {Clavel},
  {Drechsel}, {Estaria}, {Faelker}, {Riedinger}, {Robson}, {Taylor}, \&
  {Xim{\'e}nez de Ferr{\'a}n}}]{Kessler1996}
{Kessler}, M.~F., {Steinz}, J.~A., {Anderegg}, M.~E., {et~al.} 1996, \aap, 315,
  L27

\bibitem[{{Khouri} {et~al.}(2014{\natexlab{a}}){Khouri}, {de Koter}, {Decin},
  {Waters}, {Lombaert}, {Royer}, {Swinyard}, {Barlow}, {Alcolea}, {Blommaert},
  {Bujarrabal}, {Cernicharo}, {Groenewegen}, {Justtanont}, {Kerschbaum},
  {Maercker}, {Marston}, {Matsuura}, {Melnick}, {Menten}, {Olofsson},
  {Planesas}, {Polehampton}, {Posch}, {Schmidt}, {Szczerba}, {Vandenbussche},
  \& {Yates}}]{Khouri2014}
{Khouri}, T., {de Koter}, A., {Decin}, L., {et~al.} 2014{\natexlab{a}}, \aap,
  561, A5 (Paper I)

\bibitem[{{Khouri} {et~al.}(2014{\natexlab{b}}){Khouri}, {de Koter}, {Decin},
  {Waters}, {Maercker}, {Lombaert}, {Alcolea}, {Blommaert}, {Bujarrabal},
  {Groenewegen}, {Justtanont}, {Kerschbaum}, {Matsuura}, {Menten}, {Olofsson},
  {Planesas}, {Royer}, {Schmidt}, {Szczerba}, {Teyssier}, \&
  {Yates}}]{Khouri2014a}
{Khouri}, T., {de Koter}, A., {Decin}, L., {et~al.} 2014{\natexlab{b}}, \aap,
  570, A67

\bibitem[{{Koike} \& {Hasegawa}(1987)}]{Koike1987}
{Koike}, C. \& {Hasegawa}, H. 1987, \apss, 134, 361

\bibitem[{{Koike} {et~al.}(1995){Koike}, {Kaito}, {Yamamoto}, {Shibai},
  {Kimura}, \& {Suto}}]{Koike1995}
{Koike}, C., {Kaito}, C., {Yamamoto}, T., {et~al.} 1995, \icarus, 114, 203

\bibitem[{{Kozasa} \& {Sogawa}(1997{\natexlab{a}})}]{Kozasa1997a}
{Kozasa}, T. \& {Sogawa}, H. 1997{\natexlab{a}}, \apss, 255, 437

\bibitem[{{Kozasa} \& {Sogawa}(1997{\natexlab{b}})}]{Kozasa1997}
{Kozasa}, T. \& {Sogawa}, H. 1997{\natexlab{b}}, \apss, 251, 165

\bibitem[{{Levin} {et~al.}(1998){Levin}, {Gemming}, \& {Brandon}}]{Levin1998}
{Levin}, I., {Gemming}, T., \& {Brandon}, D.~G. 1998, Physica Status Solidi
  Applied Research, 166, 197

\bibitem[{{Lorenz-Martins} \& {Pompeia}(2000)}]{Lorenz-Martins2000}
{Lorenz-Martins}, S. \& {Pompeia}, L. 2000, \mnras, 315, 856

\bibitem[{{Mamon} {et~al.}(1988){Mamon}, {Glassgold}, \& {Huggins}}]{Mamon1988}
{Mamon}, G.~A., {Glassgold}, A.~E., \& {Huggins}, P.~J. 1988, \apj, 328, 797

\bibitem[{{Mathis} {et~al.}(1977){Mathis}, {Rumpl}, \&
  {Nordsieck}}]{Mathis1977}
{Mathis}, J.~S., {Rumpl}, W., \& {Nordsieck}, K.~H. 1977, \apj, 217, 425

\bibitem[{{Mattsson} {et~al.}(2007){Mattsson}, {H{\"o}fner}, \&
  {Herwig}}]{Mattsson2007}
{Mattsson}, L., {H{\"o}fner}, S., \& {Herwig}, F. 2007, \aap, 470, 339

\bibitem[{{Min} {et~al.}(2009){Min}, {Dullemond}, {Dominik}, {de Koter}, \&
  {Hovenier}}]{Min2009}
{Min}, M., {Dullemond}, C.~P., {Dominik}, C., {de Koter}, A., \& {Hovenier},
  J.~W. 2009, \aap, 497, 155

\bibitem[{{Min} {et~al.}(2003){Min}, {Hovenier}, \& {de Koter}}]{Min2003}
{Min}, M., {Hovenier}, J.~W., \& {de Koter}, A. 2003, \aap, 404, 35

\bibitem[{{Muller} {et~al.}(2008){Muller}, {Dinh-V-Trung}, {He}, \&
  {Lim}}]{Muller2008}
{Muller}, S., {Dinh-V-Trung}, {He}, J.-H., \& {Lim}, J. 2008, \apjl, 684, L33

\bibitem[{{Mutschke} {et~al.}(1998){Mutschke}, {Begemann}, {Dorschner},
  {Guertler}, {Gustafson}, {Henning}, \& {Stognienko}}]{Mutschke1998}
{Mutschke}, H., {Begemann}, B., {Dorschner}, J., {et~al.} 1998, \aap, 333, 188

\bibitem[{{Neugebauer} {et~al.}(1984){Neugebauer}, {Habing}, {van Duinen},
  {Aumann}, {Baud}, {Beichman}, {Beintema}, {Boggess}, {Clegg}, {de Jong},
  {Emerson}, {Gautier}, {Gillett}, {Harris}, {Hauser}, {Houck}, {Jennings},
  {Low}, {Marsden}, {Miley}, {Olnon}, {Pottasch}, {Raimond}, {Rowan-Robinson},
  {Soifer}, {Walker}, {Wesselius}, \& {Young}}]{Neugebauer1984}
{Neugebauer}, G., {Habing}, H.~J., {van Duinen}, R., {et~al.} 1984, \apjl, 278,
  L1

\bibitem[{{Norris} {et~al.}(2012){Norris}, {Tuthill}, {Ireland}, {Lacour},
  {Zijlstra}, {Lykou}, {Evans}, {Stewart}, \& {Bedding}}]{Norris2012}
{Norris}, B.~R.~M., {Tuthill}, P.~G., {Ireland}, M.~J., {et~al.} 2012, \nat,
  484, 220

\bibitem[{{Olofsson} {et~al.}(1996){Olofsson}, {Bergman}, {Eriksson}, \&
  {Gustafsson}}]{Olofsson1996}
{Olofsson}, H., {Bergman}, P., {Eriksson}, K., \& {Gustafsson}, B. 1996, \aap,
  311, 587

\bibitem[{{Olofsson} {et~al.}(1990){Olofsson}, {Carlstrom}, {Eriksson},
  {Gustafsson}, \& {Willson}}]{Olofsson1990}
{Olofsson}, H., {Carlstrom}, U., {Eriksson}, K., {Gustafsson}, B., \&
  {Willson}, L.~A. 1990, \aap, 230, L13

\bibitem[{{Olofsson} {et~al.}(2002){Olofsson}, {Gonz{\'a}lez Delgado},
  {Kerschbaum}, \& {Sch{\"o}ier}}]{Olofsson2002}
{Olofsson}, H., {Gonz{\'a}lez Delgado}, D., {Kerschbaum}, F., \& {Sch{\"o}ier},
  F.~L. 2002, \aap, 391, 1053

\bibitem[{{Pilbratt} {et~al.}(2010){Pilbratt}, {Riedinger}, {Passvogel},
  {Crone}, {Doyle}, {Gageur}, {Heras}, {Jewell}, {Metcalfe}, {Ott}, \&
  {Schmidt}}]{Pilbratt2010}
{Pilbratt}, G.~L., {Riedinger}, J.~R., {Passvogel}, T., {et~al.} 2010, \aap,
  518, L1

\bibitem[{{Poglitsch} {et~al.}(2010){Poglitsch}, {Waelkens}, {Geis},
  {Feuchtgruber}, {Vandenbussche}, {Rodriguez}, {Krause}, {Renotte}, {van
  Hoof}, {Saraceno}, {Cepa}, {Kerschbaum}, {Agn{\`e}se}, {Ali}, {Altieri},
  {Andreani}, {Augueres}, {Balog}, {Barl}, {Bauer}, {Belbachir}, {Benedettini},
  {Billot}, {Boulade}, {Bischof}, {Blommaert}, {Callut}, {Cara}, {Cerulli},
  {Cesarsky}, {Contursi}, {Creten}, {De Meester}, {Doublier}, {Doumayrou},
  {Duband}, {Exter}, {Genzel}, {Gillis}, {Gr{\"o}zinger}, {Henning},
  {Herreros}, {Huygen}, {Inguscio}, {Jakob}, {Jamar}, {Jean}, {de Jong},
  {Katterloher}, {Kiss}, {Klaas}, {Lemke}, {Lutz}, {Madden}, {Marquet},
  {Martignac}, {Mazy}, {Merken}, {Montfort}, {Morbidelli}, {M{\"u}ller},
  {Nielbock}, {Okumura}, {Orfei}, {Ottensamer}, {Pezzuto}, {Popesso},
  {Putzeys}, {Regibo}, {Reveret}, {Royer}, {Sauvage}, {Schreiber}, {Stegmaier},
  {Schmitt}, {Schubert}, {Sturm}, {Thiel}, {Tofani}, {Vavrek}, {Wetzstein},
  {Wieprecht}, \& {Wiezorrek}}]{Poglitsch2010}
{Poglitsch}, A., {Waelkens}, C., {Geis}, N., {et~al.} 2010, \aap, 518, L2

\bibitem[{{Posch} {et~al.}(1999){Posch}, {Kerschbaum}, {Mutschke}, {Fabian},
  {Dorschner}, \& {Hron}}]{Posch1999}
{Posch}, T., {Kerschbaum}, F., {Mutschke}, H., {et~al.} 1999, \aap, 352, 609

\bibitem[{{Sacuto} {et~al.}(2013){Sacuto}, {Ramstedt}, {H{\"o}fner},
  {Olofsson}, {Bladh}, {Eriksson}, {Aringer}, {Klotz}, \&
  {Maercker}}]{Sacuto2013}
{Sacuto}, S., {Ramstedt}, S., {H{\"o}fner}, S., {et~al.} 2013, \aap, 551, A72

\bibitem[{{Sch{\"o}ier} {et~al.}(2005){Sch{\"o}ier}, {Lindqvist}, \&
  {Olofsson}}]{Schoier2005a}
{Sch{\"o}ier}, F.~L., {Lindqvist}, M., \& {Olofsson}, H. 2005, \aap, 436, 633

\bibitem[{{Schroeder} {et~al.}(1998){Schroeder}, {Winters}, {Arndt}, \&
  {Sedlmayr}}]{Schroeder1998}
{Schroeder}, K.-P., {Winters}, J.~M., {Arndt}, T.~U., \& {Sedlmayr}, E. 1998,
  \aap, 335, L9

\bibitem[{{Sloan} {et~al.}(2003){Sloan}, {Kraemer}, {Price}, \&
  {Shipman}}]{Sloan2003}
{Sloan}, G.~C., {Kraemer}, K.~E., {Price}, S.~D., \& {Shipman}, R.~F. 2003,
  \apjs, 147, 379

\bibitem[{{Speck} {et~al.}(2000){Speck}, {Barlow}, {Sylvester}, \&
  {Hofmeister}}]{Speck2000}
{Speck}, A.~K., {Barlow}, M.~J., {Sylvester}, R.~J., \& {Hofmeister}, A.~M.
  2000, \aaps, 146, 437

\bibitem[{{Speck} {et~al.}(2011){Speck}, {Whittington}, \&
  {Hofmeister}}]{Speck2011}
{Speck}, A.~K., {Whittington}, A.~G., \& {Hofmeister}, A.~M. 2011, \apj, 740,
  93

\bibitem[{{Szymczak} {et~al.}(1998){Szymczak}, {Cohen}, \&
  {Richards}}]{Szymczak1998}
{Szymczak}, M., {Cohen}, R.~J., \& {Richards}, A.~M.~S. 1998, \mnras, 297, 1151

\bibitem[{{Takigawa} {et~al.}(2014){Takigawa}, {Tachibana}, {Huss},
  {Nagashima}, {Makide}, {Krot}, \& {Nagahara}}]{Takigawa2014}
{Takigawa}, A., {Tachibana}, S., {Huss}, G.~R., {et~al.} 2014, \gca, 124, 309

\bibitem[{{Verhoelst} {et~al.}(2006){Verhoelst}, {Decin}, {van Malderen},
  {Hony}, {Cami}, {Eriksson}, {Perrin}, {Deroo}, {Vandenbussche}, \&
  {Waters}}]{Verhoelst2006}
{Verhoelst}, T., {Decin}, L., {van Malderen}, R., {et~al.} 2006, \aap, 447, 311

\bibitem[{{Wallerstein} \& {Knapp}(1998)}]{Wallerstein1998}
{Wallerstein}, G. \& {Knapp}, G.~R. 1998, \araa, 36, 369

\bibitem[{{Woitke}(2006)}]{Woitke2006}
{Woitke}, P. 2006, \aap, 460, L9

\bibitem[{{Zeidler} {et~al.}(2013){Zeidler}, {Posch}, \&
  {Mutschke}}]{Zeidler2013}
{Zeidler}, S., {Posch}, T., \& {Mutschke}, H. 2013, \aap, 553, A81

\bibitem[{{Zhao-Geisler} {et~al.}(2011){Zhao-Geisler}, {Quirrenbach},
  {K{\"o}hler}, {Lopez}, \& {Leinert}}]{Zhao-Geisler2011}
{Zhao-Geisler}, R., {Quirrenbach}, A., {K{\"o}hler}, R., {Lopez}, B., \&
  {Leinert}, C. 2011, \aap, 530, A120

\end{thebibliography}

\begin{appendix}

\section{Silicate emission}
\label{sec:AppendixA}
\begin{table*}[B!]
\caption{Diagnostic parameters of the silicate dust species tested by us in the fitting procedure. $\kappa^{\rm max}_{9.7\,\mu{\rm m}}$, $\kappa^{\rm max}_{18\,\mu{\rm m}}$ and
$\kappa^{\rm max}_{18\,\mu{\rm m}}$/$\kappa^{\rm max}_{9.7\,\mu{\rm m}}$ represent the peak value of the opacity in the 9.7\,$\mu$m and 18\,$\mu$m silicate peaks and the ratio between these two values,
respectively. $\lambda\,\, \kappa^{\rm max}_{9.7\,\mu{\rm m}}$ and $\lambda\,\, \kappa^{\rm max}_{18\,\mu{\rm m}}$ are the wavelengths where the two peaks are found.}             % title of Table
\label{tab:optConst}      % is used to refer this table in the text
\centering                          % used for centering table
\begin{tabular}{l c c c c c c}        % centered columns (4 columns)
\hline\hline                 % inserts double horizontal lines
Species & Ref. & $\kappa^{\rm max}_{9.7\,\mu{\rm m}}$ & $\lambda\,\, \kappa^{\rm max}_{9.7\,\mu{\rm m}}$ & $\kappa^{\rm max}_{18\,\mu{\rm m}}$ & $\lambda\,\, \kappa^{\rm max}_{18\,\mu{\rm m}}$ & $\kappa^{\rm max}_{18\,\mu{\rm m}}$/$\kappa^{\rm max}_{9.7\,\mu{\rm m}}$\\    % table heading 
& & [cm$^2$/g] & [$\mu$m] & [cm$^2$/g] & [$\mu$m] & \\
\hline                        % inserts single horizontal line
Ca$_2$Al$_2$SiO$_7$ & 1 & 3210 & 10.4 & 1270 & 20.7 & 0.40 \\
Ca$_2$Mg$_{0.5}$Al$_2$Si$_{1.5}$O$_7$ & 1 & 3920 & 10.3 & 1450 & 19.4 & 0.37 \\
Mg$_{0.4}$Fe$_{0.6}$SiO$_3$ & 2 & 3050 & \ \ 9.5 & 1680 & 20.9 & 0.55 \\
Mg$_{0.5}$Fe$_{0.43}$Ca$_{0.03}$Al$_{0.04}$SiO$_3$ & 3 & 3640 &\ \ 9.7 & 1670 & 18.9 & 0.46 \\
Mg$_{0.5}$Fe$_{0.5}$SiO$_3$ & 2 & 3370 &\ \ 9.5 & 1720 & 19.3 & 0.51 \\
Mg$_{0.7}$Fe$_{0.3}$SiO$_3$ & 2 & 3940 &\ \ 9.6 & 1970 & 18.6 & 0.50 \\
Mg$_{0.7}$SiO$_{2.7}$ & 4 & 2720 &\ \ 9.3 & 1530 & 20.9 & 0.56 \\
Mg$_{0.8}$Fe$_{0.2}$SiO$_{3}$ & 2 & 4240 &\ \ 9.7 & 2000 & 18.6 & 0.47 \\
Mg$_{0.8}$Fe$_{1.2}$SiO$_{4}$ & 2 & 2610 & 10.2 & 1800 & 18.0 & 0.69 \\
Mg$_{1.5}$Fe$_{1.5}$AlSi$_{3}$O$_{10.5}$ & 1 & 3790 & 10.2 & 1700 & 18.8 & 0.45 \\
Mg$_{1.5}$SiO$_{3.5}$ & 4 & 2460 &\ \ 9.8 & 1440 & 18.9 & 0.59 \\
Mg$_{2.4}$SiO$_{4.4}$ & 4 & 1590 & 10.2 & 1370 & 17.8 & 0.86 \\
Mg$_{2}$AlSi$_{2}$O$_{7.5}$ & 1 & 4480 & 10.0 & 1740 & 18.7 & 0.39 \\
Mg$_{2}$Fe$_{2}$AlSi$_{4}$O$_{13.5}$ & 1 & 3850 & 10.1 & 1730 & 18.7 & 0.45 \\
Mg$_{2}$SiO$_{4}$ & 4 & 2440 &\ \ 9.9 & 1220 & 17.7 & 0.50 \\
Mg$_{3}$AlSi$_{3}$O$_{10.5}$ & 1 & 4390 & 10.0 & 1790 & 18.5 & 0.41 \\
Mg$_{4.5}$Fe$_{4.5}$AlSi$_{9}$O$_{28.5}$ & 1 & 3380 & 10.1 & 1820 & 18.5 & 0.54 \\
Mg$_{4}$AlSi$_{4}$O$_{13.5}$ & 1 & 4450 &\ \ 9.9 & 1810 & 18.4 & 0.41 \\
Mg$_{9}$AlSi$_{9}$O$_{28.5}$ & 1 & 4750 &\ \ 9.8 & 1980 & 18.3 & 0.42 \\
MgFeAlSi$_{2}$O$_{7.5}$ & 1 & 3640 & 10.3 & 1590 & 18.6 & 0.44 \\
MgFeSiO$_{4}$ & 2 & 2700 & 10.1 & 1873 & 18.0 & 0.69 \\
MgSiO$_{3}$ & 2 & 4910 &\ \ 9.6 & 2180 & 18.4 & 0.44 \\
\hline                                   %inserts single line
\end{tabular}
\tablefoot{ 1 - \cite{Mutschke1998}, 2 - \cite{Dorschner1995}, 3 - \cite{Jager1994}, 4 - \cite{Jager2003}}
\end{table*}

\end{appendix}

\end{document}